\documentclass[twocolumn,twocolappendix]{aastex701}
\pdfoutput=1
\usepackage{amsmath,amstext}
\let\tablenum\relax
\usepackage{siunitx}
\usepackage[T1]{fontenc}
\usepackage{times}
\usepackage{apjfonts} 
\usepackage[figure,figure*]{hypcap}
\usepackage{natbib}
\usepackage{enumerate}
\usepackage[flushleft]{threeparttable}
\usepackage{booktabs}
\usepackage{graphicx}
\usepackage{multirow}
\usepackage[encapsulated]{CJK}
\usepackage[caption=false]{subfig}
\usepackage{hyperref}
\usepackage{needspace}

\graphicspath{{./}{Figures/}}

\newcommand{\spacer}{\needspace{1\baselineskip}}
\newcommand{\grizli}{\textsc{Gri$z$li}}
\newcommand{\eazy}{\textsc{EA$z$Y}}
\newcommand{\pros}{\textsc{Prospector}}

\newcommand{\arc}[1]{\ang[angle-symbol-over-decimal]{ ; ;{#1}}}

\defcitealias{Carnall2024}{C24}

\received{2026 June 1}
\revised{2026 July 16}
\accepted{2026 August 4}
\published{2026 August 25}
\submitjournal{the Astrophysical Journal Letters}
\shorttitle{Color Gradients of Ultramassive Quiescent Galaxies}
\shortauthors{Cutler et al.}

\begin{document}
\title{Toward Unbreaking the Universe: MINERVA Measurements of Color Gradients in Massive Quiescent Galaxies Can Help Ease Too-Early Star Formation Tensions}

\correspondingauthor{Sam E. Cutler}
\email{sam.cutler@tufts.edu}

\author[0000-0002-7031-2865]{Sam E. Cutler}
\affiliation{Department of Physics and Astronomy, Tufts University, 574 Boston Avenue, Suite 304, Medford, MA 02155, USA}
\email{sam.cutler@tufts.edu}

\author[0000-0002-6265-2675]{Luke Robbins}
\affiliation{Department of Physics and Astronomy, Tufts University, 574 Boston Avenue, Suite 304, Medford, MA 02155, USA}
\email{andrew.robbins@tufts.edu}

\author[0000-0001-9002-3502]{Danilo Marchesini}
\affiliation{Department of Physics and Astronomy, Tufts University, 574 Boston Avenue, Suite 304, Medford, MA 02155, USA}
\email{danilo.marchesini@tufts.edu}

\author[0000-0002-1714-1905]{Katherine A. Suess}
\affiliation{Department for Astrophysical \& Planetary Science, University of Colorado, Boulder, CO 80309, USA}
\email{kasu8993@colorado.edu}

\author[0000-0002-9330-9108]{Adam Muzzin}
\affiliation{Department of Physics and Astronomy, York University, 4700 Keele Street, Toronto, Ontario, M3J 1P3, Canada}
\email{muzzin@yorku.ca}

\author[0000-0003-2680-005X]{Gabriel Brammer}
\affiliation{Cosmic Dawn Center (DAWN), Copenhagen, Denmark}
\affiliation{Niels Bohr Institute, University of Copenhagen, Jagtvej 128, DK-2200 Copenhagen N, Denmark}
\email{gabriel.brammer@nbi.ku.dk}

\author[0000-0003-3983-5438]{Yoshihisa Asada}
\thanks{Dunlap Fellow}
\affiliation{David A. Dunlap Department of Astronomy \& Astrophysics, University of Toronto, 50 St. George Street, Toronto, ON M5S 3H4, Canada}
\affiliation{Waseda Research Institute for Science and Engineering, Faculty of Science and Engineering, Waseda University, 3-4-1 Okubo, Shinjuku, Tokyo 169-8555, Japan}
\email{yoshi.asada@utoronto.ca}

\author[0000-0003-3243-9969]{Nicholas S. Martis}
\affiliation{University of Ljubljana, Faculty of Mathematics and Physics, Jadranska ulica 19, SI-1000 Ljubljana, Slovenia}
\email{nicholas.martis@fmf.uni-lj.si}

\author[0000-0002-8909-8782]{Stacey Alberts}
\affiliation{AURA for the European Space Agency (ESA), Space Telescope Science Institute, 3700 San Martin Drive, Baltimore, MD 21218, USA}
\email{salberts@stsci.edu}

\author[0000-0002-0243-6575]{Jacqueline Antwi-Danso}
\thanks{Banting Postdoctoral Fellow}
\affiliation{David A. Dunlap Department of Astronomy \& Astrophysics, University of Toronto, 50 St. George Street, Toronto, ON M5S 3H4, Canada}
\affiliation{Department of Astronomy, University of Massachusetts, Amherst, MA 01003, USA}
\email{j.antwidanso@utoronto.ca}

\author[0000-0001-9978-2601]{Aidan P. Cloonan}
\affiliation{Department of Astronomy, University of Massachusetts, Amherst, MA 01003, USA}
\email{apcloonan@umass.edu}

\author[0000-0002-2057-5376]{Ivo Labb\'e}
\affiliation{Centre for Astrophysics and Supercomputing, Swinburne University of Technology, Melbourne, VIC 3122, Australia}
\email{ivolabbe@gmail.com}

\author[0000-0001-8367-6265]{Tim B. Miller}
\affiliation{Center for Interdisciplinary Exploration and Research in Astrophysics (CIERA), Northwestern University, 1800 Sherman Avenue, Evanston, IL 60201, USA}
\email{timothy.miller@northwestern.edu}

\author[0000-0001-7300-9450]{Ikki Mitsuhashi}
\affiliation{Department for Astrophysical \& Planetary Science, University of Colorado, Boulder, CO 80309, USA}
\email{ikki0913astr@gmail.com}

\author[0000-0001-8592-2706]{Alexandra Pope}
\affiliation{Department of Astronomy, University of Massachusetts, Amherst, MA 01003, USA}
\email{pope@astro.umass.edu}

\author[0000-0002-1917-1200]{Anna Sajina}
\affiliation{Department of Physics and Astronomy, Tufts University, 574 Boston Avenue, Suite 304, Medford, MA 02155, USA}
\email{annie.sajina@gmail.com}

\author[0000-0001-8830-2166]{Ghassan T. E. Sarrouh}
\affiliation{Department of Physics and Astronomy, York University, 4700 Keele Street, Toronto, Ontario, M3J 1P3, Canada}
\email{gsarrouh@yorku.ca}

\author[0000-0003-1078-9706]{Monu Sharma}
\affiliation{Departament d'Astronom\'ia i Astrofis\'ica, Universitat de Val\`encia, C. Dr. Moliner 50, E-46100 Burjassot, Valencia, Spain}
\email{mnushv@gmail.com}

\author[0000-0001-7768-5309]{Mauro Stefanon}
\affiliation{Departament d'Astronom\'ia i Astrofis\'ica, Universitat de Val\`encia, C. Dr. Moliner 50, E-46100 Burjassot, Valencia, Spain}
\affiliation{Unidad Asociada CSIC ``Grupo de Astrofisica Extragal\'actica y Cosmolog\'ia'' (Instituto de F\'isica de Cantabria - Universitat de Val\`encia)}
\email{mauro.stefanon@uv.es}

\author[0009-0002-2209-4813]{Edgar P. Vidal}
\affiliation{Department of Physics and Astronomy, Tufts University, 574 Boston Avenue, Suite 304, Medford, MA 02155, USA}
\email{edgar.vidal@tufts.edu} 

\author[0000-0002-4201-7367]{Chris J. Willott}
\affiliation{National Research Council of Canada, Herzberg Astronomy \& Astrophysics Research Centre, 5071 West Saanich Road, Victoria, BC, V9E 2E7, Canada}
\email{chris.willott@nrc.ca}

\author[0000-0001-5063-8254]{Rachel Bezanson}
\affiliation{Department of Physics and Astronomy and PITT PACC, University of Pittsburgh, Pittsburgh, PA 15260, USA}
\email{rachel.bezanson@pitt.edu}

\author[0000-0001-5984-0395]{Maru\v{s}a Brada{\v c}}
\affiliation{University of Ljubljana, Faculty of Mathematics and Physics, Jadranska ulica 19, SI-1000 Ljubljana, Slovenia}
\affiliation{Department of Physics and Astronomy, University of California Davis, 1 Shields Avenue, Davis, CA 95616, USA}
\email{marusa.bradac@fmf.uni-lj.si}

\author[0000-0003-3881-1397]{Olivia R. Cooper}
\thanks{NSF Astronomy and Astrophysics Postdoctoral Fellow}
\affiliation{Department for Astrophysical \& Planetary Science, University of Colorado, Boulder, CO 80309, USA}
\email{olivia.cooper@colorado.edu}

\author[0000-0002-1109-1919]{Robert Feldmann}
\affiliation{Department of Astrophysics, Universit\"at Z\"urich, Zurich, CH-8057, Switzerland}
\email{robert.feldmann@uzh.ch}

\author[0000-0001-6003-0541]{Ben Forrest}
\affiliation{Department of Physics and Astronomy, University of California Davis, 1 Shields Avenue, Davis, CA 95616, USA}
\email{bforrest@ucdavis.edu}

\author[0000-0002-3254-9044]{Karl Glazebrook}
\affiliation{Centre for Astrophysics and Supercomputing, Swinburne University of Technology, Melbourne, VIC 3122, Australia}
\email{kglazebrook@swin.edu.au}

\author[0000-0002-5612-3427]{Jenny E. Greene}
\affiliation{Department of Astrophysical Sciences, Princeton University, 4 Ivy Lane, Princeton, NJ 08544, USA}
\email{crispygreene@gmail.com}

\author[0009-0002-5758-6025]{Valentina La Torre}
\affiliation{Department of Physics and Astronomy, Tufts University, 574 Boston Avenue, Suite 304, Medford, MA 02155, USA}
\email{valentina.la_torre@tufts.edu}

\author[0000-0002-3101-8348]{Jamie Lin}
\affiliation{Department of Physics and Astronomy, Tufts University, 574 Boston Avenue, Suite 304, Medford, MA 02155, USA}
\email{Jamie.Lin@tufts.edu}

\author[0000-0003-0695-4414]{Michael V. Maseda}
\affiliation{Department of Astronomy, University of Wisconsin-Madison, 475 N. Charter Street, Madison, WI 53706 USA}
\email{maseda@wisc.edu} 

\author[0000-0002-2446-8770]{Ian McConachie}
\affiliation{Department of Astronomy, University of Wisconsin-Madison, 475 N. Charter Street, Madison, WI 53706 USA}
\email{ian.mcconachie@wisc.edu}

\author[0000-0003-2804-0648]{Themiya Nanayakkara}
\affiliation{Centre for Astrophysics and Supercomputing, Swinburne University of Technology, Melbourne, VIC 3122, Australia}
\affiliation{Sydney Institute for Astronomy, The University of Sydney, School of Physics A28, Camperdown, 2006, NSW, Australia}
\email{themiyananayakkara@gmail.com}

\author[]{Ga\"el Noirot}
\affiliation{Space Telescope Science Institute, 3700 San Martin Drive, Baltimore, MD 21218, USA}
\email{gnoirot@stsci.edu} 

\author[0000-0002-9651-5716]{Richard Pan}
\affiliation{Department of Physics and Astronomy, Tufts University, 574 Boston Avenue, Suite 304, Medford, MA 02155, USA}
\email{richard.pan@tufts.edu}

\author[0009-0009-4283-3311]{Kesha A. Patel}
\affiliation{Department of Physics and Astronomy, Tufts University, 574 Boston Avenue, Suite 304, Medford, MA 02155, USA}
\email{kesha.patel@tufts.edu}

\author[0009-0001-8728-6894]{Veronica Pratt}
\affiliation{Department of Physics and Astronomy, Tufts University, 574 Boston Avenue, Suite 304, Medford, MA 02155, USA}
\email{veronica.pratt@tufts.edu}

\author[0000-0002-7712-7857]{Marcin Sawicki}
\affiliation{Institute for Computational Astrophysics and Department of Astronomy and Physics, Saint Mary’s University, 923 Robie Street, Halifax, Nova Scotia, B3H 3C3, Canada}
\email{marcin.sawicki@smu.ca}

\author[0000-0003-4075-7393]{David J. Setton}
\thanks{Brinson Prize Fellow}
\affiliation{Department of Astrophysical Sciences, Princeton University, 4 Ivy Lane, Princeton, NJ 08544, USA}
\email{davidsetton@princeton.edu}

\author[0000-0003-1614-196X]{John R. Weaver}
\thanks{Brinson Prize Fellow}
\affiliation{MIT Kavli Institute for Astrophysics and Space Research, 70 Vassar Street, Cambridge, MA 02139, USA}
\email{john.weaver.astro@gmail.com}

\author[0000-0002-5027-0135]{Arjen van der Wel}
\affiliation{Department of Physics and Astronomy, Ghent University, Proeftuinstraat 86, N3, B-9000 Ghent, Belgium}
\email{Arjen.vanderWel@UGent.be}

\author[0000-0001-7160-3632]{Katherine E. Whitaker}
\affiliation{Cosmic Dawn Center (DAWN), Copenhagen, Denmark}
\affiliation{Department of Astronomy, University of Massachusetts, Amherst, MA 01003, USA}
\email{kwhitaker@astro.umass.edu}

\author[0000-0001-6454-1699]{Yunchong Zhang} 
\affiliation{Department of Physics and Astronomy and PITT PACC, University of Pittsburgh, Pittsburgh, PA 15260, USA}
\email{yuz369@pitt.edu}

\author[0000-0002-1163-7790]{Kumail Zaidi}
\affiliation{Department of Physics and Astronomy, Tufts University, 574 Boston Avenue, Suite 304, Medford, MA 02155, USA}
\affiliation{HEP Division, Argonne National Laboratory, 9700 South Cass Avenue, Lemont, IL 60439, USA}
\email{kzaidi@anl.gov}

\begin{abstract}\noindent
The discovery of a population of massive, ancient quiescent galaxies within the first 2 Gyr of the Universe's history has led to significant tensions with models of galaxy formation. However, these analyses are often based on slit spectroscopy, which typically captures only the centermost region of these galaxies and, crucially, assumes these cores are representative of the entire galaxy. To illustrate the varying stellar populations present throughout these galaxies, we present an analysis of color gradients in four $z>3$, $\log(M_\star/M_\odot)>11$ quiescent galaxies which previous works have argued are in tension with models. Using medium-band photometry from MINERVA JWST observations, we measure resolved photometry in a series of elliptical annuli out to $\arc{0.7}$ ($\sim4~R_e$). We find negative color gradients in three galaxies, and for the most extreme color gradient ($\Delta(U-V)/\Delta R=-0.126\pm0.030~{\rm mag~kpc^{-1}}$), we find the stellar mass is 0.1 dex lower when compared to photometry measured within NIRSpec slits. In the limiting case where these color gradients are entirely driven by age, we find lessened tensions with extreme value statistics models out to $z\sim9.5$, though different stellar population modeling choices also contribute significantly. Ultimately, these findings highlight the need for integral field unit spectroscopy. Spatially resolved spectra can provide the evidence needed to break the age--dust--metallicity degeneracy, and reliably separate the effects of the observed color gradients from the effects of different physical modeling assumptions on the formation histories of these galaxies. 
\end{abstract}

\spacer
\section{Introduction}
Over the past decade, deep ground-based imaging and spectroscopy have revealed a population of extremely massive ($\log(M_\star/M_\odot)>11$) quiescent galaxies at $z>3$ \citep[e.g.,][]{Marsan2015,Glazebrook2017,Schreiber2018,Forrest2020,Antwi-Danso2025}. Spectroscopic observations from the James Webb Space Telescope (JWST) found that a subsample of these galaxies are very old, having formed the bulk of their stellar mass during the epoch of reionization \citep[e.g.,][]{Carnall2023a,Carnall2023b,Carnall2024,Belli2024,DEugenio2024,Glazebrook2024,Urbano2024,Xiao2024,deGraaff2025,McConachie2025,Nanayakkara2025,Hamadouche2026,Stevenson2026,Zhang2026}. The large stellar mass of these galaxies, paired with their ancient formation ages, has placed stress on cosmological models. In particular, several of these massive quiescent galaxies exceed the expected stellar mass of the most massive galaxies predicted by current models in the first $\sim1-2$ Gyr of the Universe even after assuming 100\% star formation efficiency \citep{Carnall2024,Glazebrook2024,Chittenden2026}. Several mechanisms have been proposed to explain these tensions, including a lack of understanding of early stellar populations or galaxy formation, different initial mass functions (IMFs), environmental bias \citep[e.g.,][]{Jespersen2025,Turner2025}, and challenges to $\Lambda$CDM cosmology and the cold dark matter paradigm as a whole \citep[e.g.,][]{Liu2022,Maio2023,Padmanabhan2023,Parashari2023,Sun2023,Lin2024}.

One concern is that a key assumption in these analyses has been that the spectrum observed by slits is representative of the whole galaxy. Spectroscopy from NIRSpec Micro-Shutter Assembly (MSA) shutters only covers a \arc{0.46} $\times$ \arc{0.20} area on-sky, whereas the average half-light diameter of a $3<z<4$ quiescent galaxy at $\log(M_\star/M_\odot)\sim11$ is roughly $\arc{0.3}$ \citep[e.g.,][]{Chen2026}. As such, MSA spectra of high-redshift galaxies only capture half the total light of the galaxy, and thus may be mostly dominated by light from the galaxy's center. Therefore, if we assume this inner 50\% of the galaxy's light is representative of the entire galaxy, we could be overestimating the stellar mass and age. Furthermore, inexact slit positioning (sources are rarely centered on the MSA slit) and bar shadows add further spatial complexity, which may lead to additional biases in the total galaxy values.

Lower-redshift results suggest that indeed these MSA observations may be biased by color gradients. At $z<3$, massive quiescent galaxies have strong negative color gradients (i.e., redder centers and bluer outskirts), especially in older stellar systems \citep[e.g.,][]{Guo2011,Szomoru2013,Suess2019a,Suess2020,Miller2023,Cheng2024,Clausen2025,Martorano2026}. Studies have found that in massive quiescent galaxies, color gradients are primarily driven by metallicity at $0\lesssim z\lesssim1$ \citep[e.g.,][]{Tortora2010,Greene2012,Greene2015,Cheng2024} or a mix of age and metallicity at $z\sim2$ \citep[e.g.,][]{Guo2011,Suess2021,Cheng2025}. Moreover, analyses of lensed galaxies at $z=2$ with ground-based MOSFIRE and Hubble Space Telescope (HST) grism spectroscopy have revealed largely flat age and metallicity gradients at $z\sim2$ \citep{Jafariyazani2020,Akhshik2023}, while JWST/NIRISS slitless spectroscopy found evidence for flat and positive metallicity gradients at $z\sim3$ \citep{Acharyya2025,Ju2025}. In post-starburst galaxies (PSBs), several studies find flat age gradients \citep[e.g.,][]{Setton2020,Suess2020,Suess2021}, while others find age gradients present even in PSBs \citep[e.g.,][]{DEugenio2020,Cheng2024,Leung2026}. Notably, most studies of physical parameter gradients in quiescent galaxies at $z>1$ are impacted heavily by systematics such as the spectral/spatial resolution, which emission/absorption lines are present, and contributions from diffuse ionized gas \citep[e.g.,][]{Zhang2017,Poetrodjojo2019,Ju2025}. Age gradients may even be expected in early massive quiescent galaxies due to inside out formation and quenching \citep[e.g.,][]{Kepner1999,Zolotov2015,Nelson2016,Tacchella2016,Lin2019,Pillepich2019,Spilker2019,Cutler2023} or rejuvenated star formation in outer regions due to continued gas accretion \citep[e.g.,][]{Fortune2025,Remus2025,Trussler2025}. It is thus plausible that massive quiescent galaxies at $z>3$ also exhibit strong negative color and age gradients, and subsequently younger, bluer outskirts which may ease the aforementioned tensions caused by their modeled formation timescales.

\begin{table*}[ht!]
\begin{threeparttable}
    \centering
    \caption{Summary of the Four $z>3$ Ultramassive Quiescent Galaxies with Spectroscopic Redshifts and Metallicities Reported in \citetalias{Carnall2024}.}
    \begin{tabular*}{\textwidth}{@{\extracolsep{\fill}} l c c c c c r @{}}
    \hline\hline
        ID & \citetalias{Carnall2024} ID & RA & Decl. &  Flux Radius & $z_{\rm spec}$ & $\log(Z/Z_\odot)$  \\
         & & (deg) & (deg) & & & \\\hline
        MINERVA-1084946 & ZF-UDS-6496 & 34.340393 & -5.241278 & \arc{0.22} (1.55 kpc) & 3.9884 & $0.32^{+0.04}_{-0.05}$\\
        MINERVA-1092611 & ZF-UDS-7329 & 34.255895 & -5.233866 & \arc{0.19} (1.44 kpc) & 3.1943 & $0.35^{+0.07}_{-0.08}$\\
        MINERVA-1189865 & PRIMER-EXCELS-109760 & 34.365079 & -5.148848 & \arc{0.16} (1.05 kpc) & 4.6227 & $-0.41^{+0.06}_{-0.09}$\\
        MINERVA-1208449 & PRIMER-EXCELS-117560 & 34.399674 & -5.136348 & \arc{0.17} (1.14 kpc) & 4.6194 & $0.35^{+0.08}_{-0.06}$\\
        \hline
    \end{tabular*}
    \begin{tablenotes}
        \item\hspace{-2pt}\textbf{Note.} All redshifts have uncertainties of $\pm0.0003$. The MINERVA IDs and flux radii, which are based on the n3.0\_v1.2 internal catalogs, are used throughout this work.
    \end{tablenotes}
    
    \label{tab:sample}
\end{threeparttable}
\end{table*}

Directly measuring these color gradients and obtaining more accurate stellar masses and formation ages requires spectroscopy from the NIRSpec integral field unit (IFU), deep, multiobject or long-slit spectroscopy with slits aligned along the galaxy's semi-major axis \citep[e.g.,][]{vanderWel2016,Kriek2024,Slob2024}, or NIRCam or NIRISS grism spectroscopy, all of which can account for differing stellar populations within galaxies. However, IFU observations in particular are expensive, require targeted observations for each source individually, and high-redshift sources may be too compact to justify IFU spectroscopy. 

Though JWST NIRCam imaging has enabled spatially resolved spectral energy distribution (SED) modeling, photometry from broadband imaging alone may be insufficient to obtain robust stellar population modeling. In particular, the star formation histories (SFHs) of galaxies modeled with broadband photometry alone are dominated by model priors \citep{Wang2024b}, making it difficult to constrain formation age, especially on the short timescales of the $z>3$ Universe. However, medium-band photometry, in tandem with broadbands, is capable of robustly modeling stellar populations \citep[e.g.,][]{Antwi-Danso2025}. Here, we use the ``Medium-band Imaging with NIRCam to Explore ReVolutionary Astrophysics'' (MINERVA) JWST cycle 4 treasury survey \citep[ID: 7814, PI: Muzzin, CoPIs: Marchesini, Suess;][]{Muzzin2025}, as it extensively covers the Ultra Deep Survey (UDS) field, where our target galaxies are located, and provides a minimum of 16 photometric filters (8 medium bands) when combined with existing JWST surveys. 

In this Letter, we augment existing HST and JWST data with medium-band photometry from MINERVA to perform spatially resolved SED modeling of four ultramassive ($\log(M_\odot/M_\star)>11$ quiescent galaxies at $z>3$ in the UDS field from \cite{Carnall2024}, hereafter referred to as \citetalias{Carnall2024}. The Letter is structured as follows. In Section \ref{sec:data}, we detail the MINERVA observations and data products used, as well as the four target galaxies. Section \ref{sec:methods} describes the spatially resolved photometry and three SED fitting prescriptions used to model the photometry. Color and age gradients are shown in Section \ref{sec:results}, while Section \ref{sec:discuss} discusses the impact of these results on our understanding of the $z>3$ massive quiescent galaxy population. The work is summarized in Section \ref{sec:summary}. We assume a \cite{Chabrier2003} IMF and a WMAP9 cosmology \citep{Hinshaw2013}: $H_0=69.32~{\rm km~s^{-1}~Mpc^{-1}}$, $\Omega_M=0.2865$, and $\Omega_\Lambda=0.7135$. We also assume $M_{\rm V, \odot} = +~4.81$ ABmag, taken from the CALSPEC database\footnote{\url{http://www.stsci.edu/hst/observatory/cdbs/calspec.html}}. All magnitudes are in the AB system \citep{Oke1974}.

\spacer
\section{Data}\label{sec:data}
\subsection{Imaging and Catalogs}\label{sec:catalogs}
This study utilizes internal versions of the MINERVA UDS imaging and photometric catalogs, with archival JWST and HST imaging. All MINERVA imaging and data products, including a public Data Release 1 (DR1), will ultimately be available on the MINERVA website\footnote{\url{https://minerva.colorado.edu/}} and may be obtained from the MAST archive at \dataset[doi:10.17909/cg7r-tq15]{https://doi.org/10.17909/cg7r-tq15}. Currently, a preliminary MINERVA UDS ``DR0'' is publicly available on the MINERVA website, though a full suite of papers detailing the techniques used to create the data products will not be available until DR1. For this reason, we briefly detail the general procedure behind the DR0 data products for any potential users to reference.

In addition to the MINERVA observations, we include JWST broadband imaging from the ``Public Release IMaging for Extragalactic Research'' \citep[PRIMER;][]{Dunlop2021} survey, HST imaging from CANDELS \citep{Grogin2011,Koekemoer2011}, and other overlapping public JWST imaging available within the Dawn JWST Archive\footnote{\href{https://dawn-cph.github.io/dja/}{https://dawn-cph.github.io/dja/}} (DJA). Together, MINERVA and PRIMER provide JWST/NIRCam F090W, F115W, F140M, F150W, F162M, F182M, F200W, F210M, F250M, F277W, F300M, F356W, F360M, F410M, F444W, and F460M imaging, as well as HST Advanced Camera for Surveys (ACS) F435W, F606W, and F814W and HST/WFC3 F125W, F140W, and F160W across much of the UDS field. These eight additional medium bands from MINERVA provide significantly more wavelength resolution for SED modeling, and thus are vital to robustly model the stellar populations of ancient galaxies \citep[e.g.,][]{Suess2024,Wang2024b,Antwi-Danso2025,Muzzin2025}. To ensure complete consistency with respect to the imaging reduction process across MINERVA filters and all of the additional ancillary data, we reprocess all ancillary data using the current version of the DJA imaging reduction pipeline implemented with \grizli{}  \citep{grizli}, producing new \grizli{} ``v8.0'' mosaics in all filters. 

While a complete detailing of our imaging reduction and mosaics will be provided in L. Robbins et al. (in prep.), in brief, our imaging data reduction pipeline begins with the STScI-produced \texttt{rate} files, which are derived with the STScI \textsc{Detector1Pipeline} (\texttt{calwebb\textunderscore detector1}) routine. Next, we use the \grizli{} Python package \citep{grizli}; we perform flat-fielding, astrometric alignment to previously observed JWST/NIRCam F444W public imaging, chip-level background estimation/subtraction, cosmic-ray rejection, and drizzling to a common pixel scale of $\arc{0.04}~{\rm pixel}^{-1}$ based on the \texttt{jwst\_1293.pmap} STScI Calibration Reference Data System context. To account for residual differences in the mosaic background levels (in particular around bright stars), we also perform another additional background subtraction step on the drizzled mosaics using the \textsc{Photutils} \citep{photutils2025} \texttt{Background2D} function with a box size of $12\times12$ pixels, a filter size of five boxes, and the \texttt{MedianBackground} estimator (see, e.g., Section 3.2.4 of \citealt{Sarrouh2026}). To protect extended galaxy light, we take great care to mask all objects before the spatially varying background level is estimated, and utilize the \textsc{Photutils} \texttt{BkgZoomInterpolator} method to interpolate the 2D background in masked regions. We dilate our source-masks in three ``levels'', which depend on initial estimates of the objects' isophotal fluxes such that brighter objects receive a larger dilation of their masks.

Photometric catalogs are made using an updated version of \textsc{Aperpy},\footnote{\textsc{Aperpy-2}: \url{https://github.com/astrowhit/aperpy/tree/aperpy-2}} largely following \cite{Weaver2024} and \cite{Cutler2024}, with a few significant changes. A full description of the \textsc{Aperpy-2} pipeline used for the MINERVA catalogs will be detailed in S. E. Cutler et al. (in prep.). The primary pipeline changes are to the source detection criterion. Sources are detected using a \texttt{CHI-MEAN} detection scheme \citep[e.g.,][]{Evans2001,Drlica-Wagner2018}:
\begin{align}
    \texttt{CHI-MEAN}=\frac{\sqrt{\Sigma_{c\leq n}w_cf_c^2}-\mu}{\sqrt{n-\mu^2}},
\end{align}
where $f_c$ is the background-subtracted flux of each pixel, $w_c$ is the weight of that pixel, $n$ is the number of images coadded to create the detection, and $\mu$ is the mean of a $\chi$-distribution with $n$ degrees of freedom. The detection image includes all JWST/NIRCam and HST/ACS imaging, where each image is rescaled so that the noise properties follow a $\chi$-distribution, following \citep{Sarrouh2026}. The rest of the pipeline is largely consistent with \cite{Weaver2024}. We measure catalog photometry on images convolved to the resolution of the F444W point-spread function (PSF). PSFs for each band are constructed for the photometric catalogs using stacks of unsaturated stars and homogenized to the F444W resolution with \textsc{Pypher} \citep{Boucaud2016}. Throughout this paper, all photometry and images make use of the F444W-matched MINERVA imaging. Notably, the fiducial MINERVA photometric catalogs utilize a ``super'' aperture prescription, following \cite{Labbe2003}, where aperture photometry is measured in varying aperture sizes matched to the isophotal area of the source before scaling to total photometry via the Kron radius \citep{Kron1980} and PSF curve of growth. The super apertures more accurately describe the colors of the whole galaxy compared to a fixed color aperture size, which is crucial for integrated SED modeling.

The UDS field also has significant mid-infrared photometric coverage from JWST/MIRI via PRIMER (F770W, F1800W) and MINERVA (F1280W, F1500W). MIRI imaging is independently reduced, and full details will be presented in N. S. Martis et al. (in prep.). The reduction strategy is based on that used by the SMILES survey \citep{Alberts2024}. Briefly, we run modified versions of the STScI \texttt{calwebb\_detector1} and \texttt{calwebb\_image2} pipeline steps using the \texttt{jwst\_1413.pmap} CRDS context with additional processing to account for persistence, warm pixels, and solar system objects which enter the field of view. We generate a ``super background'' by calculating the median of a stack of \texttt{cal} files with sources masked. To account for the time-varying background, visits are manually grouped for each filter in order to generate the optimal super background. A final row/column median subtraction and filtering using \textsc{Photutils} \texttt{Background2D} remove any remaining background artifacts, similar to the process for NIRCam data. We identify and mask additional persistence artifacts by performing source detection on the negative of the \texttt{stage3} \texttt{i2d} file for each visit. Finally, the MIRI astrometry is matched to the same F444W reference catalog as the NIRCam images. The MIRI mosaics use a pixel scale of \arc{0.08} (twice the size of the NIRCam mosaic pixel scale) and occupy the same footprint as the NIRCam mosaics.

Photometry in MIRI bands is measured for each source in the NIRCam-detected photometric catalog (Y. Asada et al. in prep.). For each source (with MIRI coverage) in the NIRCam catalog, we first measure the color between NIRCam/F444W and the MIRI filter of interest by performing fixed-aperture photometry on the PSF-matched NIRCam/F444W image (at MIRI resolution) and the MIRI filter image. We then use the F444W$-$MIRI color in tandem with the total F444W flux from the super aperture catalog to obtain the total flux in the MIRI filter \citep[see also, e.g.,][for a similar methodology]{Quadri2007AJ}.  

\spacer
\subsection{Target Sources}
Our analysis focuses on the four ultramassive ($\log(M_\star/M_\odot)\sim11$) quiescent galaxies in UDS analyzed with medium-resolution spectroscopy in \citetalias{Carnall2024}. Two of the galaxies (ZF-UDS-6496 and ZF-UDS-7329) were originally observed with Keck/MOSFIRE \citep{Schreiber2018} and subsequently followed up with and extensively studied prior to \citetalias{Carnall2024} using NIRSpec MSA prism spectroscopy \citep{Glazebrook2024,Nanayakkara2024}. These two galaxies also have the lowest redshifts in the sample ($z_{\rm spec}=3.9884$ and 3.1943, respectively). ZF-UDS-6496 (hereafter MINERVA-1084946) was found to have a robust PSB spectral shape. Conversely, \cite{Glazebrook2024} (and subsequently \citetalias{Carnall2024}) find ZF-UDS-7329 (MINERVA-1092611) to be extremely old, with formation redshifts consistent with $z_{\rm form}\sim11$. PRIMER-EXCELS-109760 and 117560 (MINERVA-1189865 and 1208449) are both observed at $z_{\rm spec}=4.62$, with the two galaxies possibly being part of a $z\sim4.6$ overdensity. In particular, MINERVA-1189865 is consistent with significant star formation at $z>12$. Table \ref{tab:sample} shows the four galaxies we perform follow-up analysis on, including the MINERVA ID, R.A., decl., and flux radius ($R_e$) as well as the ID, spectroscopic redshift, and metallicity from \citetalias{Carnall2024}. The MINERVA IDs are used primarily hereafter.

\spacer
\section{Analysis}\label{sec:methods}

\subsection{Resolved Photometry}\label{sec:annuli}
\begin{figure*}
    \centering
    \includegraphics[width=\linewidth]{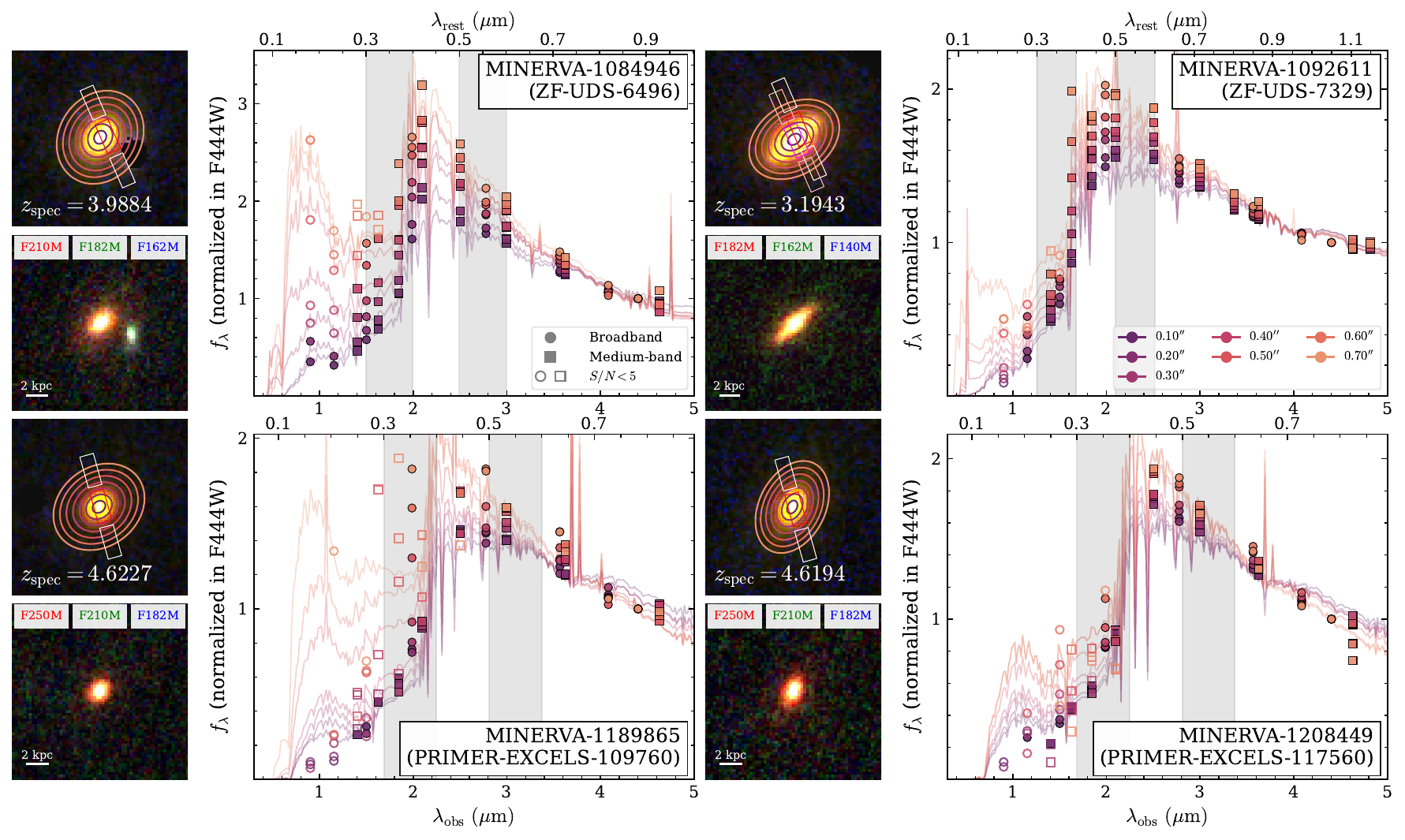}
    \caption{Image cutouts and resolved SEDs of the four ultramassive quiescent galaxies. For each galaxy, a (F090W+F115W+F150W)-F277W- F444W RGB image is shown in the top left, while the bottom left shows a medium-band-only RGB image. The three medium-band filters for each galaxy, chosen to span the Balmer/4000 \AA{} break, are indicated at the top of the cutout. A neighboring galaxy is masked in the broadband RGB cutout of MINERVA-1084946. All cutouts are made from mosaics at the native pixel scale and have a dimension of $\arc{2.5}\times\arc{2.5}$. Ellipses with colors identical to the SEDs indicate the seven annuli used to measure resolved photometry, while white/red boxes and pink/magenta boxes indicate the NIRSpec MSA slits from \citetalias{Carnall2024} and \cite{Glazebrook2024}, respectively. Resolved photometry of all NIRCam filters in each annulus is shown on the right with colored points, normalized to unity in F444W. Circles indicate archival JWST and HST data, while squares represent the MINERVA suite of medium bands. Open points indicate photometry with $S/N<5$. Thin lines show the best-fit \eazy{} model for each annulus, and gray shaded regions highlight the FWHM of the rest-frame $U$ and $V$ filters. Color gradients are clearly visible in the SEDs corresponding to different annuli.}
    \label{fig:cutouts}
\end{figure*}

We measure resolved photometry on the PSF-matched imaging in elliptical annuli using \textsc{Photutils} for all available filters, including from ancillary fields. This often results in certain sources having additional coverage compared to others (e.g., MINERVA-1092611 is the only source with F480M photometry). The outer semi-major axis of each annulus ranges from \arc{0.1} to \arc{0.7} with a width of \arc{0.1}, though for the smallest semi-major axis we consider all the flux within the \arc{0.1} elliptical aperture. Axis ratios and position angles are inherited directly from the MINERVA photometric catalog. Physically, each \arc{0.1} annulus spans between $0.67-0.77$ kpc ($0.5-0.6~R_e$) for all galaxies, depending on the redshift. The innermost annulus covers the central $\sim0.5~R_e$ and the outermost extends to between $3.2-4.4~R_e$. The outer ellipses of each annulus are shown with solid lines in the RGB images of Figure \ref{fig:cutouts} and compared to the MSA slit size and position of \citetalias{Carnall2024} (white/red slits) and \citet[pink/magenta slits]{Glazebrook2024}. Nearby sources are masked using the segmentation map for the MINERVA photometric catalogs. We do not dilate the segmentation map because the isophotal segment should cover most of the flux (see the top leftmost panel of Figure \ref{fig:cutouts}).

Photometric uncertainties are determined using the empty aperture measurements obtained during the MINERVA catalog-building process. 10,000 circular apertures are placed in empty regions of the noise-equalized, PSF-matched mosaic of each filter. The standard deviation of the distribution of empty aperture fluxes is computed and the process repeated for 30 different aperture diameters ranging from \arc{0.16} to \arc{1.4}. Following \cite{Skelton2014} and \cite{Whitaker2019}, we fit the empty aperture standard deviations with a power law of the form 
\begin{align}\label{sig}
    \sigma=\sigma_1\alpha A^{(\beta/2)},
\end{align}
where $\sigma_1$ is the pixel-to-pixel uncertainty, $A$ is the area within the aperture, and $\alpha$ and $\beta$ are free parameters. Using the best-fit values of $\alpha$ and $\beta$ and the area within each annulus, we determine the uncertainty by dividing $\sigma$ by the square root of the median weight at the location of the source. These errors reasonably account for spatial correlations caused by subtracting concentric apertures to measure annular flux. We find they fall between the case of no correlation ($\sigma_{\rm ann}^2=\sigma_{\rm out}^2+\sigma_{\rm in}^2$) and perfect correlation ($\sigma_{\rm ann}^2=\sigma_{\rm out}^2+\sigma_{\rm in}^2-2\sigma_{\rm in}\sigma_{\rm out}$).

Due to the relatively thin width of the annuli, adjacent apertures likely have significant covariance between them from PSF smoothing and a shared background model. To account for this, we repeat the empty aperture analysis; we concentrically place the seven annular apertures (including the central elliptical aperture) at each of 10,000 positions in empty regions of the mosaic. We then compute the flux covariance between aperture $i$ and $j$ via
\begin{align}
    C_{ij}=\frac{1}{N}\sum_{n=1}^{N}(f_{i,n}-\langle f_i\rangle)\times(f_{j,n}-\langle f_j\rangle),
\end{align}
where the sum is over $N=~$10,000 aperture positions. This covariance matrix is used directly when fitting observed color gradients. For rest-frame color gradients (obtained via \eazy{}, see \ref{sec:eazy}), we scale the covariance matrix by the $1\sigma$ uncertainty on the rest-frame flux in aperture $i$ and $j$ ($\sigma_i$ and $\sigma_j$):
\begin{align}
    C_{{\rm rest},ij}=(\sigma_i\times\sigma_j)\left(\frac{C_{ij}}{\sqrt{C_{ii}C_{jj}}}\right),
\end{align}
where $C_{ij}$ is the covariance of the filter whose pivot wavelength is closest to the rest-frame band.

Both the smallest aperture size and the width of each annulus exceed the NIRCam/F444W PSF half-width at half-maximum ($\sim\arc{0.075}$), which means any measured color gradients should be primarily driven by the galaxy itself. Nevertheless, as a comparison, we measure the F162M$-$F210M and F150W$-$F277W colors (PSF-matched to F444W) of point sources in the MINERVA UDS catalog in circular annuli. These colors roughly approximate the rest-frame $U-V$ at $z=3$ and 4, respectively. As expected, we find no gradients in unresolved sources, confirming that the PSF is not responsible for the observed color gradients. For simplicity, we opt not to correct our color profiles for the effects of PSF smoothing \citep[e.g.,][]{Williams2025}, though we note that the intrinsic gradient may be even steeper due to these effects. Moreover, in order to make as few assumptions as possible when measuring resolved photometry, we do not utilize any forward-modeling techniques to extract photometry using, e.g.,  S\'ersic profiles. Color gradients obtained using forward modeling are compared to our aperture method and discussed further in Appendix \ref{app:pysersic}. Overall, our primary measurements and conclusions are unchanged when using single or double S\'ersic profiles to model color gradients.

\spacer
\subsection{\eazy{}: Rest-frame Colors}\label{sec:eazy}
We utilize the photometric fitting code \eazy{} \citep{Brammer2008} to measure the rest-frame $UVJ$ colors in each annulus. While fitting, we increase the uncertainty on observed filters furthest from the rest-frame wavelength of interest (i.e., weighted interpolation) so that the rest-frame colors are better constrained by observed photometry, as opposed to using the observed photometry to correct the rest-frame colors after fitting \citep[e.g.,][]{Rudnick2003,Ilbert2005,Williams2009,Noirot2022}. Annular photometry is modeled using the \textsc{sfhz\_blue\_agn} templates with the redshift fixed at the spectroscopic redshifts shown in Table \ref{tab:sample} ($z_{\rm spec}$). For comparison, we also run \eazy{} on the super aperture and fixed \arc{0.70} color aperture photometry (both with redshifts fixed to $z_{\rm spec}$). The optional magnitude and $\beta$-slope priors are disabled in our fits, and a 5\% systematic error is applied to the photometry. 

We use \eazy{} for rest-frame colors as we find that it can more closely mimic the observed photometry. This is because \eazy{} uses a linear combination of galaxy models to match the SED, which allows for more flexibility (though note some nonstandard stellar models are not present in the \eazy{} templates; e.g., \citealt{Lu2026}), and, more importantly, determines rest-frame colors with a weighted interpolation. On the other hand, since \pros{} models are physical, we use them in Section \ref{sec:sps} for general stellar population modeling and inferring nonparametric SFHs, making the use of the two different SED modeling codes complementary.

\spacer
\subsection{\pros{}: Stellar Population Modeling}\label{sec:sps}
We model the resolved and total photometry with the Bayesian SED fitting code \pros{} \citep{prospector2021}, applying the default libraries: Flexible Stellar Population Synthesis (FSPS) models \citep{fsps2009,fsps2010a,fsps2010b}, MIST stellar isochrones \citep{mist2016a,mist2016b}, and the dust attenuation model from \citet{kreik&conroy}. We elect to use the synthetic C3K \citep{2019ConroyC3K} stellar templates due to their high rest-frame wavelength coverage ($0.01\leq\lambda\leq 2.0~\mu$m) and template stellar metallicity range ($-2.12\leq\log(Z/Z_\odot)\leq0.5$). SFHs are also modeled nonparametrically using the ``continuity'' prior with 11 total lookback time bins (two bins from 1 yr to 30 Myr and 30 Myr to 100 Myr, and 9 linearly-spaced bins from 100 Myr to $t_{\rm univ}(z=20)$). Notably, these age bins impose the condition of negligible star formation at $z>20$ following \cite{Turner2025}. This condition prevents \pros{} from ``hiding'' stellar mass at very early times and artificially inflating the mass-weighted age of the galaxy. The effect of the $z=20$ limit on the inferred SFHs is discussed in Appendix \ref{app:zmax20}. To account for any systematic uncertainties in both the observed photometry and also in our \pros{} models, we again add a 5\% systematic error floor to all photometry before running the fitting \citep{Wang2024a}. Sampling is performed using dynamic nested sampling via \textsc{dynesty} \citep{Speagle2020}. A full list of priors and prior ranges used can be found in Table \ref{table:fiducual}. Neither the prism spectroscopy from \cite{Glazebrook2024} nor the medium-resolution spectroscopy from \citetalias{Carnall2024} are included in our fitting.

The major confounding factor in our SED modeling is the age--dust--metallicity degeneracy, as all three of these parameters can contribute to reddened colors. Since the dataset used in this Letter is purely photometric in nature and spans only the rest-frame ultraviolet (UV) to near-infrared (NIR), i.e., $0.1\lesssim\lambda_{\rm rest}\lesssim1.1~\mu$ m, it is insufficient to reliably separate out our galaxies' ages, metallicities and dust attenuation parameters \citep[e.g.,][]{Bell2001,prospector2021,Tacchella2022,Nersesian2024,Nersesian2025}. We therefore chose to adopt the extreme ``limiting'' case of modeling the observed color gradients as a gradient in mass-weighted age by imposing narrow priors on the dust and metallicity parameters (see Table \ref{table:fiducual}). These ``fiducial'' priors are designed to test whether such maximized age gradients \textit{could} help resolve the tensions our galaxies' SFHs pose to theoretical models. Ultimately, IFU spectroscopy is required to determine which physical properties are driving the color gradients for each galaxy individually.

We first constrain the stellar metallicity in each annulus with a narrow Gaussian prior based on the \citetalias{Carnall2024} best-fit metallicities and uncertainties, with the caveat that \citetalias{Carnall2024} fits use different stellar modeling \citep[\textsc{Bagpipes},][]{Carnall2018,Carnall2019} and different assumptions about the dust and SFH. Thus, by construction we implicitly assume no strong metallicity gradients in these galaxies, which may disagree with studies of lower-redshift massive quiescent galaxies. However, at $z\sim3.5$, \cite{Edward2026} find evidence for largely flat [Fe/H] gradients in massive PSBs with positive age gradients. Moreover, they also find negative [$\alpha$/Fe] gradients in two of their three galaxies, in contrast to the flat [$\alpha$/Fe] gradients observed at low redshifts, suggesting the assumption of flat metallicity gradients may not be unrealistic given the short timescales involved at $z>3$.

To better constrain the dust attenuation, we model the super aperture total photometry, including MIRI photometry for all galaxies except MINERVA-1092611, which has no MIRI coverage. Total mass formed and dust attenuation ($\tau_V$) are fit with uniform priors ranging from $6<\log(M_{\rm form}/M_\odot)<12.5$ and $0<\tau_V<4$, respectively. The $\tau_V$ from the super photometry run is then used as a strong prior ($\sigma_{\tau_V}=0.046$) for fits to the annular photometry. A full description of the priors, including the values we use as constraints on the dust and metallicity priors used in our fiducial \pros{} modeling, is detailed in Appendix \ref{app:priors}. The impact of this flat dust gradient assumption on our age gradients is likely less severe than a metallicity gradient. While a gradient in dust attenuation may soften any observed age gradients, birth cloud dust could alternatively lead to preferential attenuation of O/B type stars, which would make any observed age gradients stronger in reality \citep[e.g.,][]{Salim2020}. 

\begin{figure*}[ht!]
    \centering
    \includegraphics[width=\linewidth]{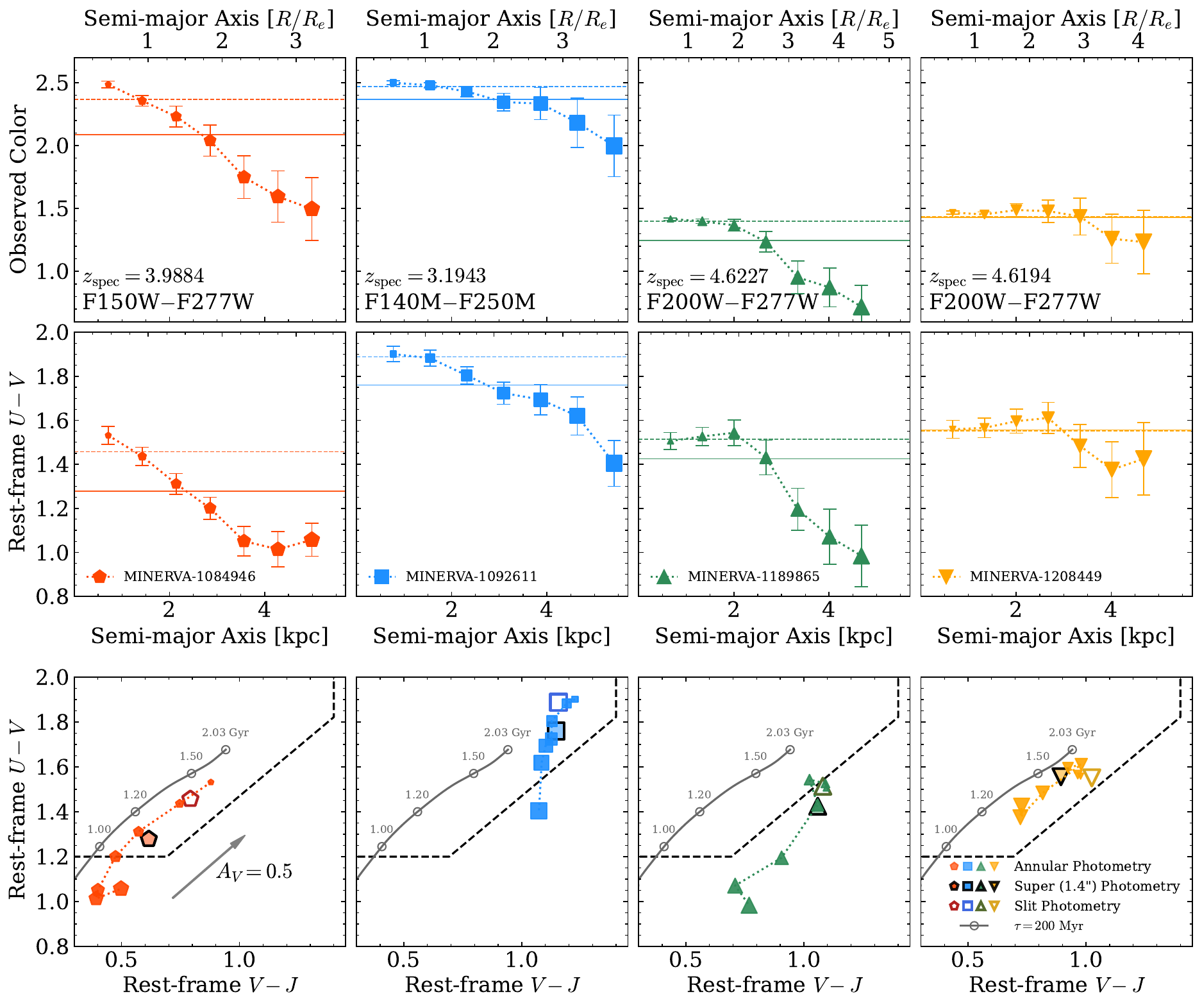}
    \caption{Most of the ultramassive quiescent galaxies at $z>3$ have negative color gradients. Top row: change in observed colors with aperture semi-major axis (in kiloparsecs on the bottom axis and relative to the flux radius on the top) for each galaxy in our sample. The filters, chosen to be closest to the pivot wavelength of the rest-frame $U$ and $V$ bands, are indicated in the bottom left of each panel. Filled points (connected via dotted lines) indicate the color in each annulus, with larger-sized points indicating larger radii. Solid lines indicate colors from super aperture photometry (see Section \ref{sec:data}), while dashed lines show colors derived from ``slit photometry'' (photometry measured in rectangular apertures approximating the NIRSpec/G235M slits of \citetalias{Carnall2024}) with \eazy{}. Note, both the observed and rest-frame colors of the super and slit photometry are roughly identical for MINERVA-1208449. Middle row: rest-frame $U-V$ color gradients. Symbols are identical to the top row. Bottom row: evolution of annular rest-frame colors within the $UVJ$ diagram. Colored points are identical to the top row. Open points indicate rest-frame colors from slit photometry, while black-outlined points represent colors from super apertures. Apertures closer in size to the isophotal area of the galaxy (e.g., the \arc{1.4} super aperture) are more representative of the entire galaxy, while smaller apertures (e.g., \arc{0.32}) can be strongly biased by light from the galactic center. Black dashed lines indicate the $UVJ$ selection for quiescent galaxies at $2<z<3.5$ from \cite{Whitaker2011}. Dark gray solid lines indicate the rest-frame color evolution of an FSPS model with an exponentially-declining SFH that has an $e$-folding timescale of $\tau=200$ Myr.}
    \label{fig:uvj}
\end{figure*}

Notably, we assume a Chabrier IMF \citep{Chabrier2003} for both global and annular SED fits. This assumption is likely simplistic, as several studies have found evidence for top-heavy IMFs at $z>10$ \citep[e.g.,][]{Hutter2025, Jeong2025} and bottom-heavy IMFs in the central $\sim1$ kpc of $z<1$ massive galaxies \citep[e.g.,][]{vanDokkum2010,Smith2020,Cheng2026}. In the  former scenario, we would instead measure lower stellar masses, due to the lower mass-to-light ratios caused by the overabundance of massive stars, which could lessen tensions with galaxy formation models alone. Conversely, using bottom-light IMFs in fits to the central annuli would further stress these models, as noted in \cite{Cheng2026}. However, with the data at hand, we cannot directly disentangle which IMFs are present in these galaxies. Moreover, rather than assume a different IMF, we instead use a Chabrier IMF throughout to show how considering the resolved populations within $z>3$ massive quiescent galaxies (along with other sensible modeling assumptions) can begin to ease tensions with models alone.

\spacer
\section{Results}\label{sec:results}
\subsection{Observed Color Gradients}
Figure \ref{fig:cutouts} shows the resolved SEDs (normalized to unity in F444W) of the ultramassive quiescent galaxies from \citetalias{Carnall2024}, alongside (F090W+F115W+F150W)-F277W-F444W and medium-band-only RGB cutouts of the galaxies. Qualitatively, the SEDs are clearly different, with the large scatter in the rest-optical ($0.4\lesssim\lambda_{rest}\lesssim0.6~\mu$ m) illustrating the varying spectral shapes. The effect is stronger in MINERVA-1084946 and MINERVA-1092611 due to their lower redshifts, which allow for higher signal-to-noise-ratio ($S/N$) photometry at shorter wavelengths out to larger radii.

In MINERVA-1084946, we observe bluer UV slopes at $R\geqslant\arc{0.4}$ relative to smaller radii, though at $R\geqslant\arc{0.6}$ this is mostly driven by low-$S/N$ photometry ($<5$) at $\lambda_{\rm rest}<0.3$ in these annuli. The SEDs at $R\geqslant\arc{0.4}$ also demonstrate a steeper slope between $0.4\lesssim\lambda_{\rm rest}\lesssim0.6~\mu$m compared to those in central regions. Meanwhile, in the outer annuli ($R\geqslant\arc{0.4}$) of MINERVA-1092611 the SED sharply decreases from 2 to 3 $\mu m$, while it is much flatter in the galactic center. These bluer rest-optical SEDs may indicate an age gradient \citep[though a strong dust gradient is also a potential explanation; e.g.,][]{Siegel2025}. In MINERVA-1092611, owing to the combination of F140M, F150W, and F162M photometry in MINERVA, we are also able to distinguish a stronger Balmer break ($\lambda_{\rm obs}=1.53~\mu$m) at larger radii, compared to a more pronounced 4000 \AA{} ($\lambda_{\rm obs}=1.68~\mu$m) break in the inner radii. This transition from a prominent Balmer break in the outer annuli to a 4000 \AA{} break in the galactic center suggests a younger stellar population in the galactic outskirts. 

The top row of Figure \ref{fig:uvj} shows observed color gradients for each galaxy. We choose the broadband filters (to maximize $S/N$) that best approximate a rest-frame $U-V$ color, though we opt to use F140M$-$F250M for MINERVA-1092611 as its lower redshift enables high-$S/N$ observations with medium bands. MINERVA-1084946 and 1189865 have $2.9\sigma$ and $1.4\sigma$ significant negative color gradients ($-0.218\pm0.075$ and $-0.090\pm0.065~{\rm mag~kpc^{-1}}$), respectively. However, MINERVA-1084946 has a steeper color gradient in the center ($R\lesssim3$ kpc, \arc{0.5}), while 1189865 is steeper at $R\gtrsim2$ kpc (\arc{0.3}). If we only consider the colors in these regimes, the gradients become significantly more prominent ($-0.192\pm0.010$ and $-0.275\pm0.066~{\rm mag~kpc^{-1}}$, respectively). Conversely, we find that MINERVA-1208449 has a flat gradient ($-0.022\pm0.071~{\rm mag~kpc^{-1}}$). Likewise, MINERVA-1092611 is consistent with a flat color gradient ($-0.054\pm0.065~{\rm mag~kpc^{-1}}$), though if we only consider the central $\sim2.5$ kpc (\arc{0.4}) where $S/N$ is higher, we find a mild, statistically significant negative color gradient ($-0.044\pm0.005~{\rm mag~kpc^{-1}}$). The flatter gradient in MINERVA-1092611 (both centrally and globally) compared to the steeper rest-frame $U-V$ color gradient we find in Section \ref{sec:rfcolor} is likely due to its unfortunate redshift ($z_{\rm spec}=3.1943$), which places the rest-frame $V$-band in the wavelength gap between the short- and long-wavelength NIRCam channels and makes it difficult to find a NIRCam filter that is a close $V$ band proxy, as well as artificial flattening due to PSF smoothing (see Section \ref{sec:annuli}).

\spacer
\subsection{Rest-Frame Colors}\label{sec:rfcolor}
In the middle panel of Figure \ref{fig:uvj}, we show the rest-frame $U-V$ color gradients derived from \eazy{}. For all galaxies except MINERVA-1208449, we find apparent negative color gradients in at least some region of the galaxy. Quantitatively, MINERVA-1084946 has the strongest color gradient ($-0.131\pm0.018~{\rm mag~kpc^{-1}}$), though MINERVA-1189865 also has a $>3\sigma$ negative gradient ($-0.123\pm0.039~{\rm mag~kpc^{-1}}$). Interestingly, MINERVA-1092611 has a $>3\sigma$ significant rest-frame $U-V$ color gradient ($-0.084\pm0.017~{\rm mag~kpc^{-1}}$), while its observed F140M-F250M color gradient was less significant, especially across all annuli. This is likely due to \eazy{} leveraging multiple filters to measure the rest-frame fluxes.  

The bottom panel of Figure \ref{fig:uvj} shows the evolution of the $UVJ$ colors of each galaxy with annular distance. At larger radii (indicated by larger points), galaxies tend to have dramatically bluer rest-frame colors, even when compared to the super aperture total photometry (black-outlined points). Notably, the outer annuli colors of all galaxies except MINERVA-1208449 fall below the \cite{Whitaker2011} $2<z<3.5$ quiescent galaxy selection (dotted black line, left), though these are still classified as quiescent when using high-redshift-specific $UVJ$ selections \citep[e.g.,][]{Baker2025}. MINERVA-1208449 becomes $\sim0.2$ mag bluer in both $U-V$ and $V-J$ color, while MINERVA-1084946 and 1189865 show even larger ($>0.4$ mag) decreases in both $U-V$ (though these colors may be model-dependent due to high photometric uncertainty at $\lambda_{\rm rest}<0.4~\mu$m) and $V-J$ colors. In particular, the color evolution of MINERVA-1084946 and 1189865 also follows the slope of the \cite{Whitaker2011} diagonal $UVJ$ cut (dashed line in Figure \ref{fig:uvj}, bottom left). This relation is known to roughly trace the aging of quiescent galaxies due to the transition from Balmer-break-dominated to 4000 \AA{}-break-dominated spectra \citep{Whitaker2012,Whitaker2013,Belli2019}, implying that these two galaxies' color gradients are due to age (though they may also be driven by dust as shown by the $A_V$ vector in Figure \ref{fig:uvj}). Lastly, MINERVA-1092611 has the most distinct color gradient from the other three galaxies, with a sharp ($>0.4$ mag) drop in $U-V$, but relatively little change in $V-J$. We note that the $V-J$ colors are extrapolated for all galaxies except MINERVA-1092611 due to a lack of rest-frame $J$-band coverage, though we do find rough agreement between the annular $V-J$ colors for these galaxies and the global fits which have rest-$J$ coverage thanks to MIRI.

The negative rest-frame color gradients seen in three of the galaxies suggest that stellar populations in the outskirts of these ultramassive quiescent galaxies are younger. Recently, several studies \citep{Turner2025,Kawinwanichakij2026} explored the semiresolved, rest-frame colors of MINERVA-1092611 via a bulge--disk decomposition and single S\'ersic profile fitting and found a disk color consistent with an $0.1-0.5$ Gyr simple stellar population, compared to $3-5$ Gyr in the bulge. This bulge age estimate is older than the age of the Universe at $z\sim3.2$ ($t_{\rm obs}\sim 1.97$ Gyr), which suggests that this galaxy has a significant amount of dust attenuation leading to its redder colors, or that its stellar population is not well represented by a coeval simple stellar population (i.e., a more complex SFH is needed). Notably, we find that the galaxy is far redder in outer annuli than the \cite{Turner2025} disk measurements. This is likely driven by two factors: differences between \eazy{} and \pros{} rest-frame colors and the addition of medium-band photometry. \cite{Turner2025} infer $UVJ$ colors by correcting observed F150W, F200W, and F444W magnitudes to rest-frame $U$, $V$, and $J$ via the best-fit \pros{} model, which has more restricted colors than \eazy{} (see Section \ref{sec:sps}). The addition of medium bands also allows for more robust determination of rest-frame fluxes, especially in the $V$ band where Balmer and 4000 \AA{} breaks are the dominant spectral features.

\begin{figure*}[p!]
    \centering
    \includegraphics[width=0.95\linewidth]{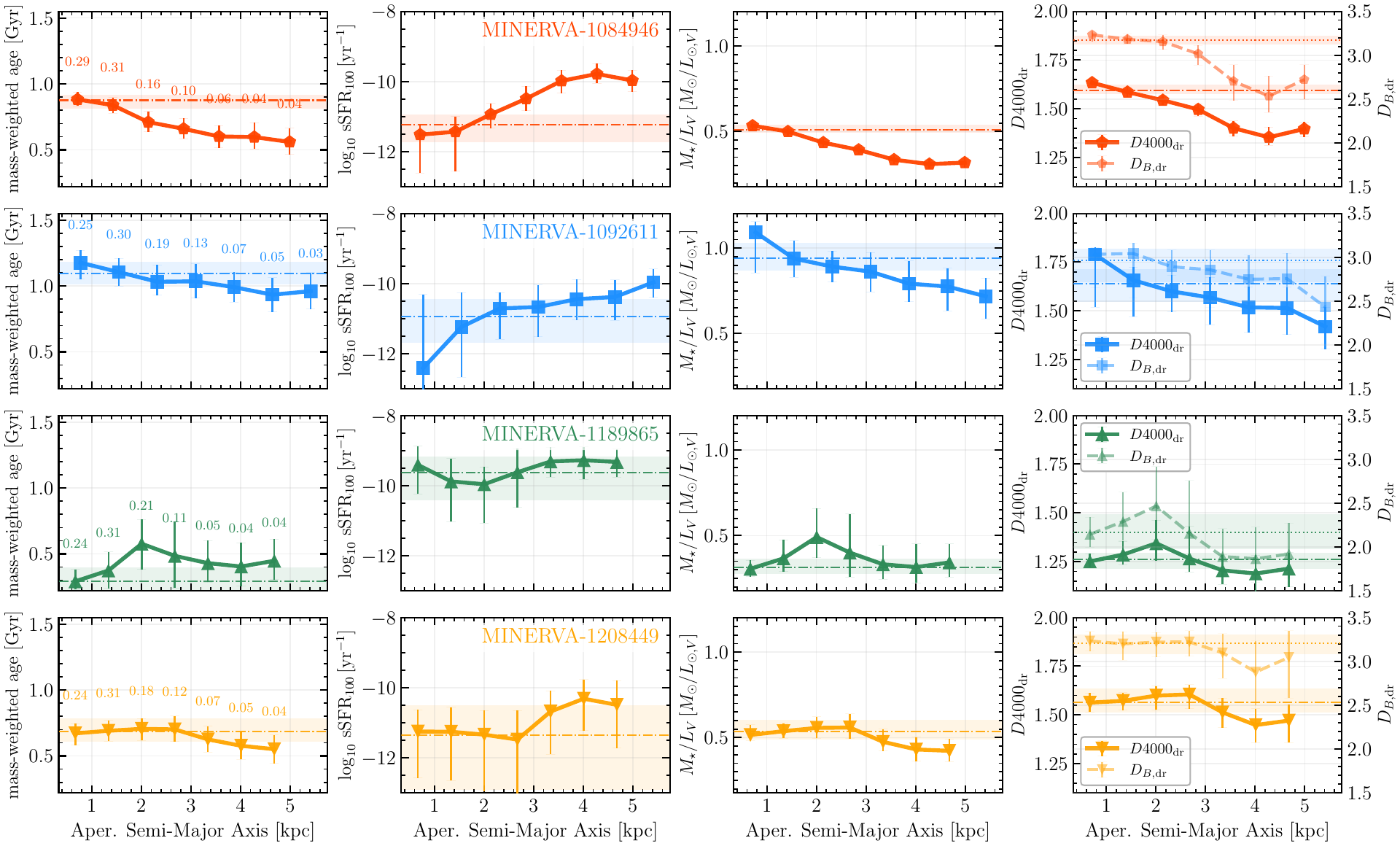}
    \label{fig:params}
    \caption{The distribution of galaxy mass-weighted age, sSFR (averaged over the last $100$ Myr of the galaxies' SFHs), rest-frame $V$-band mass-to-light ratio, and dust-corrected (``de-reddened'') $D4000$ and $D_B$ ($D4000_{\rm dr}$ and $D_{B, \rm dr}$) strengths in each annular aperture for our ``fiducial'' model, whereby the metallicity is given a strong Gaussian prior set by the values from \citetalias{Carnall2024}, and a strong dust prior set by inferred global values from modeling including MIRI photometry. The fractions above the values in the mass-weighted age column indicate the relative amount of the total stellar mass contained within the corresponding annulus. Error bars indicate $1\sigma$ uncertainties. Dashed--dotted (dotted) lines and lightly shaded regions indicate the median mass-weighted age, sSFR, $M_\star/L_{\rm V}$, and $D_{4000,{\rm dr}}$ ($D_{B,{\rm dr}}$) and their uncertainties as measured from fits to the slit photometry using our fiducial model.}

    \vspace{0.5cm}
    \centering
    \includegraphics[width=0.9\linewidth]{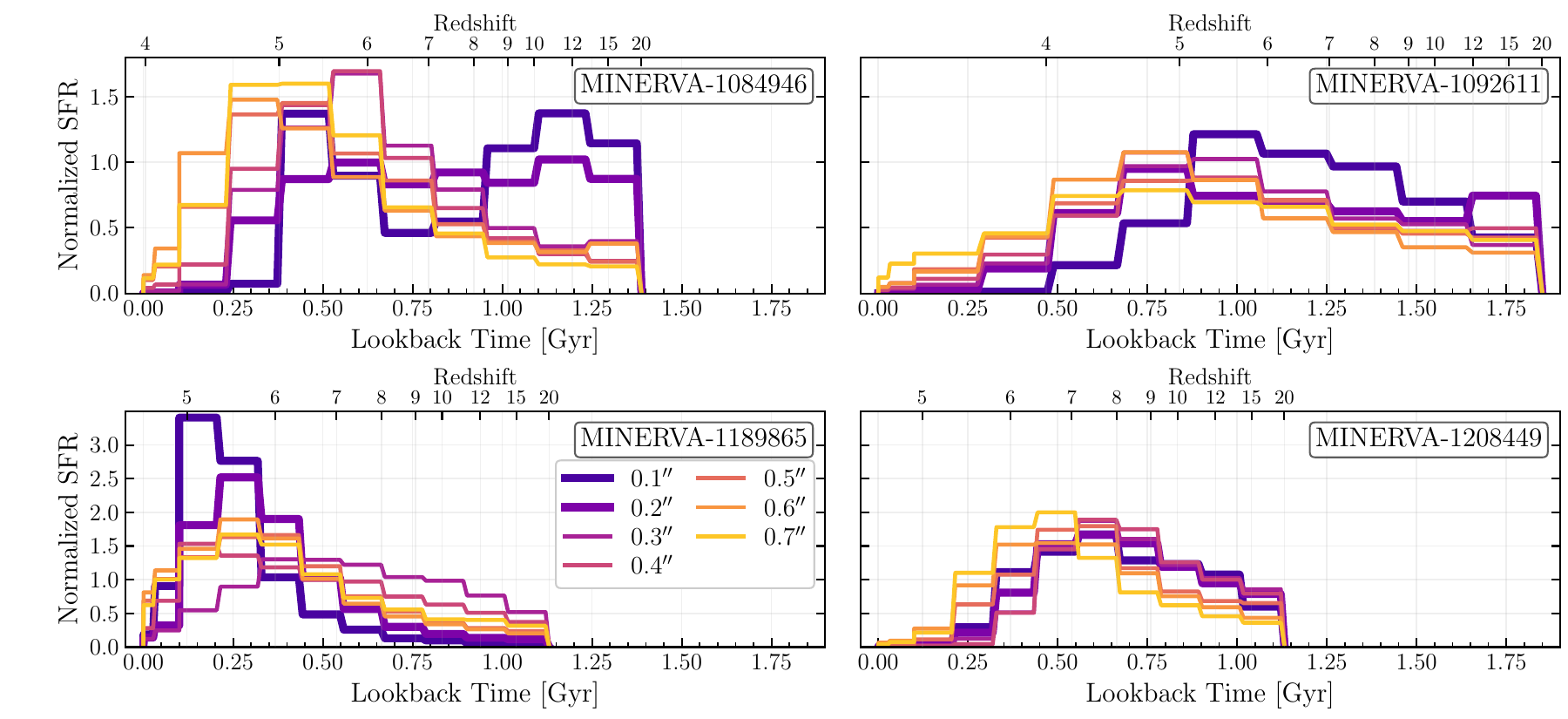}
    \caption{Resolved, median SFHs from \pros{} for the four \citetalias{Carnall2024} ultramassive quiescent galaxies. Colors indicate the corresponding annulus that was used to measure photometry and are identical to Figure \ref{fig:cutouts}. SFHs are normalized such that their integral is unity. In all galaxies except MINERVA-1189865, stellar populations in the outskirts form the bulk of their stars later on.}\label{fig:sfhs}

\end{figure*}

The outer $UVJ$ colors of MINERVA-1084946 and 1208449 (Figure \ref{fig:uvj}, red and yellow points, respectively) extend into the PSB region of the $UVJ$ diagram \citep[e.g.,][]{Belli2019,Park2023}. In particular, FSPS models suggest that the resolved colors of MINERVA-1084946 and 1208449 occupy a similar parameter space to an exponentially declining SFH model with an $e$-folding time of 200 Myr (gray lines in Figure \ref{fig:uvj}). This agrees with the findings of \citetalias{Carnall2024}, who measure PSB-like SFHs for both these galaxies. MINERVA-1189865 follows a similar trajectory to the FSPS models but is significantly bluer overall. This is also reflected in the \pros{} results (Figure \ref{fig:params}), which show a young ($<500$ Myr) population at almost all radii and a higher specific star formation rate (sSFR) than all other galaxies in the sample. Conversely, the color gradient of MINERVA-1092611 follows an older, slow-quenching color evolution, in line with the findings of \cite{Glazebrook2024}.

Crucially, colors measured from total photometry (solid or dashed lines and open or black-outlined points in Figure \ref{fig:uvj}) are strongly biased by light from the galactic center. If photometry is measured in rectangular apertures approximating the central NIRSpec/G235M slitet of \citetalias{Carnall2024} (``slit photometry''; open points), we find observed and rest-frame colors in line with the \arc{0.2} annuli. Larger \arc{0.7} diameter apertures (such as those used in the \citealt{Glazebrook2024} photometry) also have colors roughly equivalent to smaller diameter annuli (e.g., \arc{0.3}) and redder than the photometry at larger radii. This suggests that studies utilizing small color apertures or slit photometry \citep[e.g.,][]{Nanayakkara2025,Zhang2026} are often biased significantly by the central population of the galaxy, which could lead to biases in stellar population modeling \citep[see also][]{Pacifici2015}. Alternatively, the super photometry (black-outlined points), which uses \arc{1.4} diameter apertures for all galaxies herein, is comparable to the \arc{0.4} annular photometry. Thus, in the absence of resolved photometry, choosing an aperture size closest to the isophotal area is key to measuring colors representative of the entire galaxy, especially in galaxies with strong color gradients. 

\spacer
\vspace{-0.1cm}
\subsection{Stellar Population Modeling}
Figure \ref{fig:params} shows the distribution of \pros{}-measured mass-weighted age, sSFR, and strengths of the de-reddened (dust-corrected) $D4000$ \citep{Bruzual1983, Poggianti1997} and Balmer breaks \citep[$D_B=f_\nu(4160-4290$ \AA{}$)/f_\nu(3500-3630$ \AA{}$)$,][]{Roberts-Borsani2024} as a function of aperture size for our fiducial model, while Figure \ref{fig:sfhs} shows the median \pros{}-modeled SFHs for each individual annulus. In order to explore the effect of a maximal age gradient, metallicity and dust gradients are flat by construction (see Section \ref{sec:sps}), though are shown in Appendix \ref{app:priors}. In general, under this assumption, relatively strong color gradients in two of the galaxies lead to age and sSFR gradients in our stellar population modeling. We also find consistently steep declines in the de-reddened $D4000$ and $D_B$ strengths with increasing aperture size in those same galaxies. Figure \ref{fig:params} also shows the change in rest-frame $V$ band mass-to-light ratio ($M_\star/L_V$) with increasing radial distance. The mass-to-light ratio follows the same trend as the mass-weighted age, with the centers of MINERVA-1084946 and 1092611 containing far more mass per unit luminosity than the outskirts at fixed IMF. Moreover, SFHs mostly indicate that central populations have enhanced star formation at earlier times (e.g., $z>8$ in MINERVA-1084946 and $z>5$ in MINERVA-1092611), while the SFHs in outer annuli peak later ($z<6$ and $z<5$, respectively). 

The two lowest-redshift galaxies, MINERVA-1084946 and 1092611, show negative age gradients, decreasing by $0.32\pm0.12$ and $0.22\pm0.17$ Gyr, respectively, between the inner and outermost annuli. These two galaxies also have up to 1 dex higher sSFRs in their outskirts, while the de-reddened $D4000$ and $D_B$ strength decrease by 0.2 and 0.5, respectively, at larger radii. MINERVA-1084946 shows a very early burst of star formation in the central \arc{0.2}, with later bursts and rapid quenching in the outer regions. Comparatively, MINERVA-1092611 has a central SFH that drops quickly in the center, but remains elevated in the outskirts until later times, which contributes to the negative age gradients we observe. 

MINERVA-1208449 is consistent with both a flat age and sSFR gradient within the $1\sigma$ uncertainty. Interestingly, the 4000 \AA{} break in MINERVA-1208449 shows a significant decrease beyond $R=\arc{0.4}$, but a much smaller change in Balmer break strength, though this may be an effect of low-$S/N$ short-wavelength ($\lambda_{\rm obs}<2~\mu$m) photometry in the outermost annuli. In MINERVA-1208449, the photometry in the  $R>\arc{0.4}$ annuli have $S/N<3$ at $\lambda_{\rm rest}\lesssim0.4~\mu$m (see Figure \ref{fig:cutouts}, bottom right), which makes it difficult to constrain the stellar population models at these wavelengths. This impacts the ability to measure the Balmer break strength ($3565-4225$ \AA{}) more than $D4000$ ($3750-4250$ \AA{}).

MINERVA-1189865 has a largely flat age gradient, with all annuli at $R>\arc{0.3}$ having an age of roughly 0.5 Gyr, though the innermost annulus is significantly younger ($0.29^{+0.09}_{-0.06}$ Gyr). In fact, this is the youngest galaxy overall, which is consistent with the bluer rest-frame colors in Figure \ref{fig:uvj}. The flatter age and sSFR gradients may also be due to the more compact size of the galaxy, even compared to the other $z\sim4.6$ galaxy (MINERVA-1208449). Regardless, the $D4000$ and $D_B$ indices show significant drop-offs at $R>\arc{0.3}$, though the break strengths are notably low. Likewise, the sSFRs are high ($\log({\rm sSFR}/{\rm yr}^{-1})>-9.4$), only $\sim0.5$ dex below the \cite{Popesso2023} star-forming main sequence (SFMS). Together, these physical properties suggest that MINERVA-1189865 is a PSB, which is discussed further in Section \ref{sec:psb}.

\begin{table*}[ht!]
\begin{threeparttable}
    \centering
    \caption{Total Stellar Masses and Mass-weighted Ages for each Galaxy}
    \label{table:mass}
    \begin{tabular*}{\linewidth}{@{\extracolsep{\fill}}lcccccr@{}}
        \hline\hline
        \multirow{2}{*}{ID} & \multicolumn{3}{c}{Total $\log(M_\star/M_\odot)$} & \multicolumn{3}{c}{Mass-weighted Age (Gyr)} \\
        \cline{2-4}\cline{5-7}
         & Annular Sum & Slit & Super (w/MIRI) & Annular Weighted Average & Slit & Super (w/MIRI) \\
        \hline
        1084946 & $10.97^{+0.01}_{-0.01}$ & $11.07^{+0.02}_{-0.02}$ & $11.02^{+0.01}_{-0.01}$ & $0.78^{+0.03}_{-0.03}$ & $0.88^{+0.04}_{-0.07}$ & $0.78^{+0.07}_{-0.09}$ \\
        1092611 & $11.16^{+0.02}_{-0.03}$ & $11.19^{+0.04}_{-0.03}$ & $11.15^{+0.07}_{-0.06}$ & $1.08^{+0.05}_{-0.05}$ & $1.09^{+0.09}_{-0.09}$ & $1.08^{+0.19}_{-0.22}$ \\ 
        1189865 & $10.76^{+0.06}_{-0.05}$ & $10.73^{+0.08}_{-0.07}$ & $10.60^{+0.16}_{-0.06}$ & $0.42^{+0.07}_{-0.07}$ & $0.29^{+0.10}_{-0.07}$ & $0.27^{+0.22}_{-0.10}$ \\
        1208449 & $11.01^{+0.02}_{-0.02}$ & $11.06^{+0.05}_{-0.04}$ & $11.07^{+0.03}_{-0.04}$ & $0.68^{+0.04}_{-0.04}$ & $0.69^{+0.10}_{-0.08}$ & $0.70^{+0.15}_{-0.19}$ \\
        \hline
    \end{tabular*}
    \begin{tablenotes}
        \item\hspace{-2pt}\textbf{Note.} Stellar masses (ages) are calculated by summing the annular masses (taking a weighted average of the ages), scaling the slit photometry to total, or using the super photometry from the MINERVA catalogs. As the annular photometry only includes flux out to $R=\arc{0.7}$, the slit and super masses are decreased by a factor of 1.12 to remove the PSF growth curve scaling applied to account for light at $R>\arc{0.7}$.
    \end{tablenotes}
\end{threeparttable}
\end{table*}

We also test whether \pros{} interprets the strong color gradients as age gradients, even without strong priors on attenuation and metallicity. In models where the resolved SEDs are fit with a uniform prior on $A_V$, we find that \pros{} prefers a moderate age gradient combined with a sharp decrease in dust at larger radii. Likewise, models where the best-fit metallicities from \citetalias{Carnall2024} are not used as priors find very strong metallicity gradients, though these are likely nonphysical as the jump in metallicity exceeds 1 dex between some annuli and the median metallicities across all annuli are inconsistent with the measurements from \citetalias{Carnall2024}. However, if real, these strong metallicity gradients ($\Delta\log(Z/Z_\odot)/\Delta\log(R/{\rm kpc})<-2$) are comparable to predictions of monolithic collapse simulations \citep[e.g.,][]{Kobayashi1999}, which has implications for $z>3$ quiescent galaxy assembly. Regardless, while the observed color gradients are likely driven by a mix of age, dust, and metallicity, this Letter is exploring the extreme case of maximal age gradients to test if these color gradients could explain any discrepancies with cosmological models.

Table \ref{table:mass} shows the total mass of each galaxy computed via a sum of the annular masses compared to the slit and super photometry. We find that the $\log(M_\star/M_\odot)$ for the total galaxy is slightly lower (though typically within the $1\sigma$ uncertainty) when computed with a resolved approach compared to those from the slit photometry in all galaxies except MINERVA-1189865, which has a very young center. In the galaxies with negative rest-frame color and age gradients (MINERVA-1084946 and 1092611), we find that the super photometry is comparable to the annular sum. Similarly, in Table \ref{table:mass}, we compare the mass-weighted age computed via a weighted average of all annular bins to those from the slit and super photometry. The resolved approach finds that MINERVA-1084946 is younger than the slit values by $>3\sigma$ and MINERVA-1189865 is older by roughly the same margin, though the other two galaxies are consistent with the slit and super fits. Overall, the impact of color gradients on the total mass and age of ultramassive quiescent galaxies varies between galaxies. For strong negative gradients (e.g., MINERVA-1084946), we find that slit photometry can indeed overestimate the mass-weighted age and stellar mass of a galaxy. However, the reverse is true for MINERVA-1189865, which globally has a negative color gradient, though in the central 2 kpc the color gradients are essentially flat while the age gradient is positive. Furthermore, the manner of stellar population modeling (i.e., code, dust/metallicity treatment, SFH prior) can significantly change derived physical parameters regardless of the way photometry is handled.

\spacer
\section{Discussion}\label{sec:discuss}
We find evidence of negative color gradients in two of the four UDS ultramassive quiescent galaxies from \citetalias{Carnall2024}. In the test case of flat dust and metallicity gradients, we also find  negative (MINERVA-1084946 and 1092611), flat (MINERVA-1208449 and 1189865 at $R\geqslant\arc{0.3}$), and mildly positive (MINERVA-1189865 at $R\leqslant\arc{0.3}$) age gradients via \pros{} SED modeling. Irrespective of the stellar population modeling, the existence of strong color gradients in high-redshift ultramassive quiescent galaxies implies that single-slit spectroscopic observations do not always capture the full extent of the spatial variations found within these galaxies. Thus, scaling the mass and other physical properties from the single slit, which covers only half the light of the galaxy if perfectly centered, to the total flux of the galaxy can result in biased stellar masses and global ages and exacerbate tensions with existing models of galaxy evolution.

Similarly, in the case of maximal age gradients, we find resolved SFHs that suggest older or even entirely different stellar populations when compared to the \citetalias{Carnall2024} SFHs derived from slit-based spectroscopy, though the differences in stellar population modeling also play a significant role. In the following section, we highlight how analyzing these ultramassive quiescent galaxies in a resolved context with our fiducial model can help ease tensions with models -- further emphasizing the need for resolved IFU spectroscopy.

\begin{figure*}
    \centering
    \includegraphics[width=\linewidth]{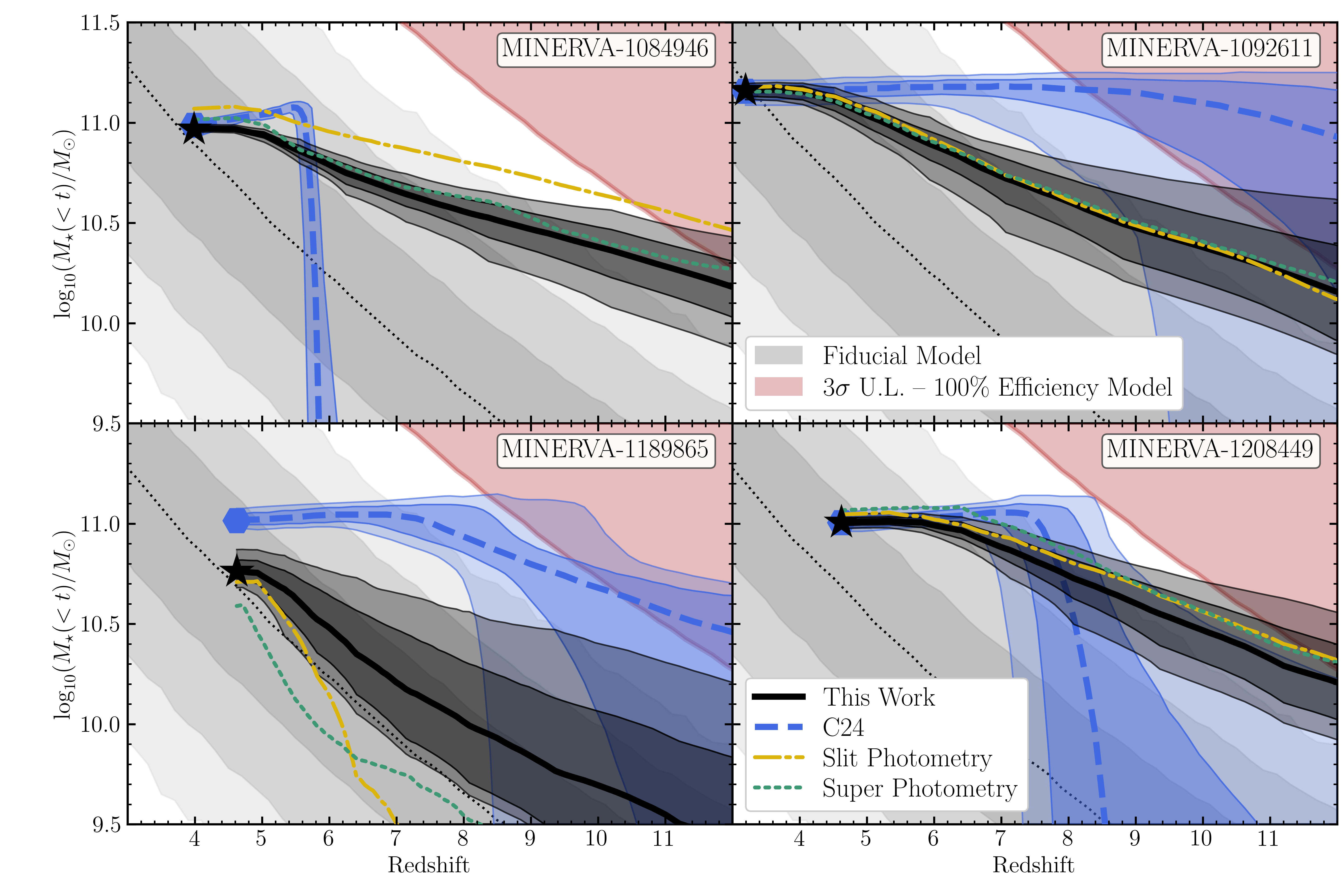}
    \caption{Stellar mass assembly histories of the four ultramassive quiescent galaxies based on our fiducial \pros{} model. The sum of the median stellar mass assembly histories (defined as the cumulative sum of the mass of living stars and stellar remnants) for each of the annuli are shown with thick black lines. $1$ and $2\sigma$ uncertainties are determined by a quadrature sum of the uncertainties on the assembly histories of the 7 individual annuli and are indicated by black shading. Median stellar mass assembly histories of the slit photometry are shown as the green dashed line, and median histories using the super+MIRI photometry used to inform our strong dust prior are shown as the gold dash--dotted line (see Appendices \ref{app:priors} and \ref{app:miri-seds}). Slit and super+MIRI assembly histories are decreased by a constant factor of 1.12 (see note in Table \ref{table:mass}). Dashed curves show the stellar mass assembly histories from \citetalias{Carnall2024}, with $1\sigma$ and $2\sigma$ confidence intervals shaded. MINERVA-1084946 only shows $1\sigma$ uncertainties as the $2\sigma$ intervals are not shown in \citetalias{Carnall2024}. The $1\sigma$, $2\sigma$, and 3$\sigma$ regions of the \cite{lovell23} EVS based stellar mass limits are shown as a function of redshift for the \cite{lovell23} fiducial model (determined with dark matter halo efficiencies drawn from a truncated log-normal distribution where $\mu=e^{-2}$ and $\sigma=1$, i.e., $0.050<f_\star<0.368$) as the gray contours. Stellar masses which would exceed the the upper 3$\sigma$ limit determined by the \cite{lovell23} model with 100$\%$ halo efficiency are denoted by the shaded red region. The specific EVS models shown here also assume an on-sky area of 160 arcmin$^2$, to match the selection function and analysis used in \citetalias{Carnall2024}. Our \pros{} modeling of resolved photometry finds much less tension with the fiducial EVS model when compared to the \citetalias{Carnall2024} assembly histories.}
    \label{fig:evs}
\end{figure*}

\spacer
\subsection{Inferred Stellar Mass Assembly Histories}
High-redshift, ultramassive quiescent galaxies pose a problem for existing models of galaxy formation due to the large amount of stellar mass they build up at early times. \citetalias{Carnall2024} consider this tension with an extreme value statistics (EVS) approach \citep{Harrison2011,lovell23}. The EVS framing predicts the stellar mass of the most massive expected galaxy within the volume of the PRIMER (and MINERVA) UDS observations. With this analysis, \citetalias{Carnall2024} find that MINERVA-1092611 and 1189865 both exceed the $3\sigma$ upper limit of the EVS model, predominantly at $z>8$. Notably, \citetalias{Carnall2024} ignore potential impacts from mergers. While \citetalias{Carnall2024} disregard major mergers due to the short time intervals over which these galaxies formed, minor mergers could also contribute to the overall stellar mass and build the strong color gradients we observe. At lower redshifts ($z<3$), color gradients are likely (re)established by the ``inside-out'' growth of quiescent galaxies; minor mergers deposit younger/metal-poor stars at large radii, leading to bluer colors in the outskirts \citep{Bezanson2009,Naab2009,Suess2019a}. However, it is unclear if there is sufficient time at $z>3$ for minor mergers to create a color gradient in the $<1$ Gyr post-quenching. Studies place timescales for minor mergers at $\tau_e\geqslant1$ Gyr \citep[e.g.,][]{Newman2012}, with the fastest occurring on $\tau_e\sim1$ Gyr \citep{Suess2019b}. For that reason, we also consider the growth of these galaxies in a merger-free context, though we note that merger rates may be faster in more overdense environments and at higher redshifts \citep[e.g.,][]{Rodriguez-Gomez2015,Duan2025,Puskas2025}, and thus may contribute significantly to the formation of massive quiescent galaxies at $z>3$ \citep{Cochrane2025}.

In Figure \ref{fig:evs}, we consider the \citetalias{Carnall2024} ultramassive quiescent galaxies in the EVS framework using our resolved SED modeling approach. For comparison, we use the same fiducial EVS model as \citetalias{Carnall2024}, which consists of a truncated log-normal distribution of baryon conversion efficiency ($f_\star$) with mean $\mu=e^{-2}$ and standard deviation $\sigma=1$ (i.e., $0.050<f_\star<0.368$). We then compute the stellar mass assembly histories for each annulus individually using the best-fit SFHs and summed into a mass assembly history for the total galaxy. While the total stellar masses remain large relative to the most massive expected galaxy, we find less tension overall with the fiducial EVS model. In particular, the median mass assembly histories only exceeds the $3\sigma$ upper limit at $z\gtrsim11$ in MINERVA-1084946 and MINERVA-1092611 and $z\gtrsim10$ in MINERVA-1208449. MINERVA-1189865 is also consistent with the fiducial EVS model within $2\sigma$. Crucially, the best-fit assembly histories never exceed the $3\sigma$ limit of an EVS with fixed $f_\star=1$ (i.e., all baryons are converted into stars) at $z<12$ (i.e., $t_{\rm lookback}<0.95-1.65$ Gyr). Beyond these redshifts, our nonparametric SFHs are less certain, in part due to larger redshift bins and fewer observational constraints, which is shown by the Bayes' factor test in Appendix \ref{app:zmax20}. Thus, while we do still find that the mass assembly histories of MINERVA-1092611 and 1208449 exceed the $f_\star=1$ EVS model at $z>12$, any tensions at these redshifts are likely entirely driven by the choice of prior for the SFH.

The mass assembly histories inferred from \citetalias{Carnall2024} are shown with blue dashed lines in Figure \ref{fig:evs}. These curves largely imply much earlier formation and higher overall stellar masses than our resolved approach. MINERVA-1092611 and 1189865 have higher median stellar mass at all epochs, with the former exceeding 100\% halo efficiencies as late as $z\sim8.5$. Notably, \citetalias{Carnall2024} argue that the lower $2\sigma$ limit of the SFH is more informative than the median/best-fit model, primarily because current SFH models are biased toward older ages \citep[e.g.,][]{Carnall2018,Leja2019a,Leja2019b}. Indeed, the $2\sigma$ lower limits for all four \citetalias{Carnall2024} mass assembly histories fall within the $3\sigma$ scatter of the fiducial EVS model. However, for MINERVA-1092611 and 1189865 these limits still find significantly larger masses than even our median assembly histories at $z\lesssim8.5$, further emphasizing the impact of both resolved photometry and SED modeling choices.

\citetalias{Carnall2024} also find that MINERVA-1084946 and 1208449 have significantly different mass assembly histories than our analysis. For both galaxies, the SFHs from \citetalias{Carnall2024} are more PSB-like and closer in shape to the outer annuli modeled herein (Figure \ref{fig:sfhs}, top left). In particular, for MINERVA-1084946, \citetalias{Carnall2024} also find a significantly younger mass-weighted age ($0.55\pm0.03$ Gyr) compared to our resolved approach ($0.78\pm0.07$ Gyr). This discrepancy in age and SFH is driven by three factors: differences in the SFH model (\citetalias{Carnall2024} use a parameterized double power law), differences in measured dust attenuation, and the \citetalias{Carnall2024} slits being offset from the center of MINERVA-1084946 (see Figure \ref{fig:cutouts}, top left). In particular, \citetalias{Carnall2024} find $A_V=0.49\pm0.05$ compared to our value of $0.14\pm0.02$, which is based on global fits including MIRI photometry. When we fit the annular photometry with an uninformative prior on $A_V$ we find a much higher attenuation in the center ($0.57\pm0.07$). In global fits without MIRI photometry, we find that $A_V$ remains close to the best-fit value with MIRI.

There are several differences between the SED modeling analysis of \citetalias{Carnall2024} and this work, including different codes (\textsc{Bagpipes} vs \pros{}), SFH assumptions (double power law parametric versus nonparametric continuity prior), and dust attenuation models (\citealt{Salim2018} versus \citealt{kreik&conroy}). Clearly, the differences in the mass assembly histories we find relative to \citetalias{Carnall2024} are in part due to these different modeling assumptions and techniques. To determine qualitatively how much of this is due to modeling (compared to color gradients), in Figure \ref{fig:evs} we also include the median mass assembly histories from \pros{} fits to the slit photometry (gold dashed--dotted lines) and the super+MIRI photometry (green dashed lines). In all four galaxies, we find that our \pros{} mass assembly histories differ significantly from the \citetalias{Carnall2024} measurements. However, for MINERVA-1084946 and 1189865, we also find that the mass assembly histories measured with slit photometry differ from our resolved approach. These two galaxies have the most significant observed and rest-frame color gradients (and subsequently age gradients as well), which explains the different assembly histories. For MINERVA-1084946, the strong negative color and age gradients cause the slit photometry to overestimate the stellar mass growth and age (see also Table \ref{table:mass}). Meanwhile, the positive age gradient in the central $\sim2$ kpc of MINERVA-1189865 causes the reverse problem; the young central bins dominate the integrated SED from the slit (and even super) photometry, leading to underestimated ages and masses.

For both MINERVA-1092611 and 1208449, we find consistent mass assembly histories in all three runs (summed annuli, slit, and super+MIRI). This is reasonable, as we find flat to mildly negative observed color and age gradients in these galaxies. Thus, any differences with \citetalias{Carnall2024} (and \citealt{Glazebrook2024} for MINERVA-1092611) are primarily due to different SED modeling choices. In particular, since \cite{Glazebrook2024} also use \pros{} with a continuity prior on the SFH to fit MINERVA-1092611, differences between our models are likely due to the imposed limit on star formation prior to $z=20$ and the fact that they limit fitting to $\lambda_{\rm rest}<0.7~\mu$m. Rest-NIR data are critical for constraining the age, stellar mass, and SFH of a galaxy \citep[e.g.,][]{Papovich2001,Papovich2023,Pforr2012, Conroy2013,Mobasher2015,Harvey2025}, and as such our use of the C3K stellar templates (which allow us to fit to $\lambda_{\rm rest}\sim1.1~\mu$m for MINERVA-1092611) may lead to different median values for these parameters. We also note that the \citetalias{Carnall2024} slit for MINERVA-1092611 (white/red boxes in Figure \ref{fig:cutouts}), which is used to compute the slit photometry, is slightly off center compared to the \cite{Glazebrook2024} slit (pink/magenta boxes). Crucially, the \citetalias{Carnall2024} slit misses the central bin, which is oldest and has the highest $M_\star/L_V$. This could further contribute to the final mass assembly histories being closer to the super and annular runs, whereas photometry from a centrally aligned slit would find a different, elevated mass assembly history closer to \cite{Glazebrook2024} and \citetalias{Carnall2024}.

Our results imply that both strong color gradients in ultramassive quiescent galaxies, particularly in MINERVA-1084946, and different modeling prescriptions could alleviate tensions with galaxy formation and evolution models. While significant contributions from dust or metallicity gradients could lessen age gradients and increase tension with early formation times, this study effectively serves as a test case for maximal age gradients. Moreover, when dust attenuation is left as a free parameter in \pros{} SED fits (i.e., only metallicity is constrained to the \citetalias{Carnall2024} values), we still find lessened tension with models compared to the mass assembly histories from \citetalias{Carnall2024}. This is likely due to the use of a nonparametric SFH prior that prevents mass from being formed at $z>20$. While this prescription only prevents star formation in the first $\sim180$ Myr of the Universe, any significant star formation at these redshifts causes extreme tension with the EVS models in Figure \ref{fig:evs}, as the most massive halo would only contain $M_\star\sim10^{8}~M_\odot$ at $z=20$. The effect of the $z=20$ SFH prior is detailed further in Appendix \ref{app:zmax20}.

\spacer
\subsection{MINERVA-1189865: A possible PSB at $z=4.62$?}\label{sec:psb}
MINERVA-1189865 differs significantly from the other ultramassive quiescent galaxies in this sample. Resolved \pros{} fits find mass-weighted ages consistently $\lesssim500$ Myr, with the innermost annuli having ages $<400$ Myr. Notably, the SFH in the central bin is indicative of a PSB, with $90\%$ of the stellar mass forming in the $\sim430$ Myr between $z=8$ and $5.4$ before quenching over the next $\sim230$ Myr, as shown by the dark blue line in the bottom-left panel of Figure \ref{fig:evs}. The sSFR gradients support a similar PSB nature, as the $\log({\rm sSFR}/{\rm yr}^{-1})\sim-9.4$ measured in the central \arc{0.1} occupies a ``transitional'' region, being roughly 0.5 dex below the \cite{Popesso2023} SFMS and indicating a very recent quenching event \citep[e.g.,][]{Tacchella2019}. Similarly, the de-reddened $D4000$ and $D_B$ strengths of the $R\leq\arc{0.4}$ annuli in MINERVA-1189865 are firmly in the PSB range \citep[$D_B>2.0$ and $D4000<1.4$, e.g.,][]{Poggianti1997,Roberts-Borsani2024,Steinhardt2024,Wilkins2024}, further suggesting the central regions of this galaxy have recently and rapidly quenched. The \citetalias{Carnall2024} stellar population modeling finds $t_{\rm form}=0.80\pm0.05$ Gyr for MINERVA-1189865, over twice as old as our best-fit age of the central $R=\arc{0.1}$ aperture ($0.29\pm0.07$ Gyr). The older global formation age inferred by \citetalias{Carnall2024} is likely due to the older ages in the $R=\arc{0.2}$ and \arc{0.3} annuli, which contribute $\sim50\%$ of the total mass of MINERVA-1189865 and thus dominate the photometry at $R<\arc{0.3}$, as well as the higher dust attenuation ($A_V=0.84\pm0.09$ in \citetalias{Carnall2024} compared to $0.74\pm0.03$ herein). If the annular photometry for MINERVA-1189865 is modeled with an uninformative prior on $A_V$, we find a more comparable global SFH to \citetalias{Carnall2024}, though the central aperture still retains a PSB-like SFH irrespective of the dust prior, further illustrating the importance of a resolved analysis. 

Significantly, the SFHs and mass-weighted stellar ages of outer annuli ($\arc{0.3}\geqslant R$) in MINERVA-1189865 indicate an older stellar population than the center. This positive age gradient in the central \arc{0.3} may be a result of a recent secondary starburst in the galactic center \citep[e.g.,][]{Dekel2014,Zolotov2015,Tacchella2018,Nelson2019}. The lack of high levels of star formation prior to the recent starburst (see the blue line in the bottom-left panel of Figure \ref{fig:evs}) may be primarily due to ``outshining'' \citep{Papovich2001}, in which rest-optical emission---MINERVA photometry only covers out to rest-frame $\sim0.85~\mu$m at $z=4.62$---is dominated by light from younger stars, thus making it challenging to disentangle previous episodes of star-formation from the most recent burst in the absence of rest-frame NIR photometry. Several recent studies utilizing JWST/NIRSpec IFU spectroscopy have also found similar flat/slightly-positive age gradients and young galactic centers in $\log(M_\star/M_\odot)>11$ PSBs at $z>3$ \citep{DEugenio2024,Edward2026}. These results highlight that massive quiescent galaxies can have a variety of gradients in their colors and physical parameters, even at $z>3$. Regardless of whether the observed gradients are positive, negative, or flat, fixed-aperture photometry and/or slit spectroscopy are insufficient to determine the nature of the whole galaxy, therefore producing biased star formation and assembly histories.

\spacer
\section{Summary}\label{sec:summary}
In this work, we examine the colors and physical properties of four ultramassive, $z>3$ quiescent galaxies presented in \citetalias{Carnall2024} and \cite{Glazebrook2024} within a resolved framework. Utilizing high-quality mosaics (L. Robbins et al. in prep.; N. S. Martis et al. in prep.) and photometric catalogs (S. E. Cutler et al. in prep.; Y. Asada et al. in prep.) of archival broadband and MINERVA medium-band data, we measure the observed SEDs of these four galaxies in a series of \arc{0.1} elliptical annuli. The resolved photometry is modeled with \eazy{} and \pros{} to measure gradients in rest-frame $UVJ$ colors and stellar population parameters. In our stellar population modeling, the dust attenuation and stellar metallicity are constrained using global measurements (including MIRI photometry) and medium-resolution spectroscopy from \citetalias{Carnall2024}, respectively, in order to maximize any potential age gradients. Our main results are listed below:

\begin{enumerate}[(i)]
    \item Ultramassive quiescent galaxies at $z>3$ can have strong gradients in both observed and rest-frame color. In the extreme case of MINERVA-1084946, we find an observed F150W$-$F277W color gradient of $-0.218\pm0.075~{\rm mag~kpc^{-1}}$ and a rest-frame $U-V$ color gradient of $-0.126\pm0.030~{\rm mag~kpc^{-1}}$. These redder centers can lead to overestimated stellar masses and ages when photometry is measured in small apertures or slits and scaled to the total flux of the galaxy. As a result, these ultramassive galaxies should be examined for any spatial gradients before inferring global stellar population properties.
    
    \item Stellar population modeling from \pros{} suggests that there could be significant negative and positive age gradients in MINERVA-1084946 and the central 2 kpc of MINERVA-1189865, respectively, with more mild gradients in the other two galaxies (MINERVA-1092611 and 1208449). However, in reality, dust and metallicity gradients are also possible, which would temper these maximized age gradients. Follow-up IFU spectroscopy is \textit{critical} to reveal which physical parameters drive the observed color gradients.

    \item Combining the mass assembly histories of each spatial bin, we find less tension with the EVS models of \cite{lovell23}, especially at $z<9$. Both resolved galaxy structure (i.e., color gradients, especially in MINERVA-1084946) and stellar population modeling systematics tangibly affect measured mass assembly histories early in the Universe. At higher redshifts, SFHs are much more uncertain, especially in the absence of resolved rest-NIR data \citep{Papovich2023}.  
\end{enumerate}

These results suggest that global ages and stellar masses of compact ($R_e\sim1-1.5$ kpc) massive quiescent galaxies at $z>3$ should not be inferred from a single small photometric aperture without first checking for spatial gradients. Likewise, careful consideration of the stellar population modeling and priors is crucial, as both resolved population structure and modeling systematics can influence stellar population measurements and inferred mass assembly histories. Compared to the elliptical morphologies dominating massive quiescent galaxies at lower redshifts, there is evidence that some of these galaxies have both disky and elliptical substructures \citep[e.g.,][]{Turner2025,Kawinwanichakij2026}, which suggests at least two distinct stellar populations within the galaxy, even in the absence of mergers. It is therefore possible that at $z>3$, the stellar masses and ages of massive quiescent galaxies can be overestimated when using global photometry in tandem with single-slit spectroscopy, particularly in cases where the slit is not aligned to the semi-major axis of a galaxy. Moving forward, deeper photometry at bluer wavelengths ($\lambda_{\rm rest}<4000$ \AA{}) is crucial in improving constraints on rest-frame colors and ages, especially in the faint galactic outskirts. Ultimately, while our photometric analysis provides evidence of these effects, IFU spectroscopy will be needed to fully break the age--dust--metallicity degeneracy in a resolved framework.

\begin{acknowledgments}
This work is based in part on observations made with the NASA/ESA/CSA James Webb Space Telescope obtained from the Space Telescope Science Institute, which is operated by the Association of Universities for Research in Astronomy, Inc., under NASA contract NAS 5–26555. Financial support for program JWST-GO-7814 is gratefully acknowledged and is provided by NASA through grants from the Space Telescope Science Institute, which is operated by the Association of Universities for Research in Astronomy, Incorporated, under NASA contract NAS 5-03127. 

The authors acknowledge the Tufts University High Performance Computing Cluster,\footnote{\url{https://it.tufts.edu/high-performance-computing}} which was utilized for the research reported in this paper. This research used the Canadian Advanced Network For Astronomy Research (CANFAR) operated in partnership by the Canadian Astronomy Data Centre and The Digital Research Alliance of Canada with support from the National Research Council of Canada, the Canadian Space Agency, CANARIE, and the Canadian Foundation for Innovation. The Cosmic Dawn Center is funded by the Danish National Research Foundation under grant DNRF140.

D.M. acknowledges generous support from the Leonard and Jane Holmes Bernstein Professorship in Evolutionary Science. A.M. acknowledges the support of the Canadian Space Agency (CSA) via grant 25JWGO4A08, as well as support from the Yavin Family Fund. M.B., N.M. acknowledge support from the ERC Grant FIRSTLIGHT \#101053208, Slovenian national research agency ARIS through grants N1-0238 and P1-0188. T.B.M. was supported by a CIERA Fellowship. M.S.T. and M.S.H. acknowledge support from the European Research Commission Consolidator Grant 101088789 (SFEER). M.S.T. also acknowledges support from the CIDEGENT/2021/059 grant by Generalitat Valenciana, and from project PID2023-149420NB-I00 funded by MICIU/AEI/10.13039/501100011033 and by ERDF/EU. O.R.C. acknowledges that support for this work was provided by the National Science Foundation Astronomy \& Astrophysics Postdoctoral Fellowship Award No. 2503202. B.F. acknowledges support from JWST-GO-02913. M.V.M. was supported by NASA via STScI grant JWST-AR-04948. D.J.S. and J.R.W. acknowledge that support for this work was provided by The Brinson Foundation through a Brinson Prize Fellowship grant.  
\end{acknowledgments}

\facilities{JWST (NIRCam and MIRI), HST (WFC3 and ACS).}

\software{
\textsc{Aperpy} \citep[\url{github.com/astrowhit/aperpy}]{Weaver2024},
\textsc{astrodrizzle} \citep{Gonzaga2012},
\textsc{Astropy} \citep{astropy2013,astropy2018,astropy2022},
\textsc{DYNESTY} \citep{Speagle2020},
\eazy{} \citep{Brammer2008},
\textsc{extinction} \citep{extinction},
\textsc{FSPS} \citep{fsps2009,fsps2010a,fsps2010b},
\grizli{} \citep[\url{github.com/gbrammer/grizli}]{grizli},
\textsc{matplotlib} \citep{matplotlib2007},
\textsc{numpy} \citep{numpy2011},
\textsc{Photutils} \citep{photutils2025},
\pros{} \citep{prospector2021},
\textsc{Pypher} \citep{Boucaud2016},
\textsc{Pysersic} \citep{Pasha2023},
\textsc{Python-FSPS} \citep{pythonfsps2014},
\textsc{SciPy} \citep{scipy2020}
\textsc{SEP} \citep{Barbary2016},
\textsc{SFDMap} \citep[\url{github.com/kbarbary/sfdmap}]{Schlegel1998,Schlafly2011}
}

\setcounter{section}{0}
\renewcommand{\thesection}{\Alph{section}}
\renewcommand{\thesubsection}{\thesection.\arabic{subsection}}
\renewcommand{\thefigure}{\Alph{section}\arabic{figure}}
\renewcommand{\thetable}{\Alph{section}\arabic{table}}
\section*{Appendix}

\section{Comparing Aperture- and Model-based Approaches to Resolved Photometry} \label{app:pysersic}
\setcounter{figure}{0}
\setcounter{table}{0}

\begin{figure*}[t!]
    \centering
    \includegraphics[width=\linewidth]{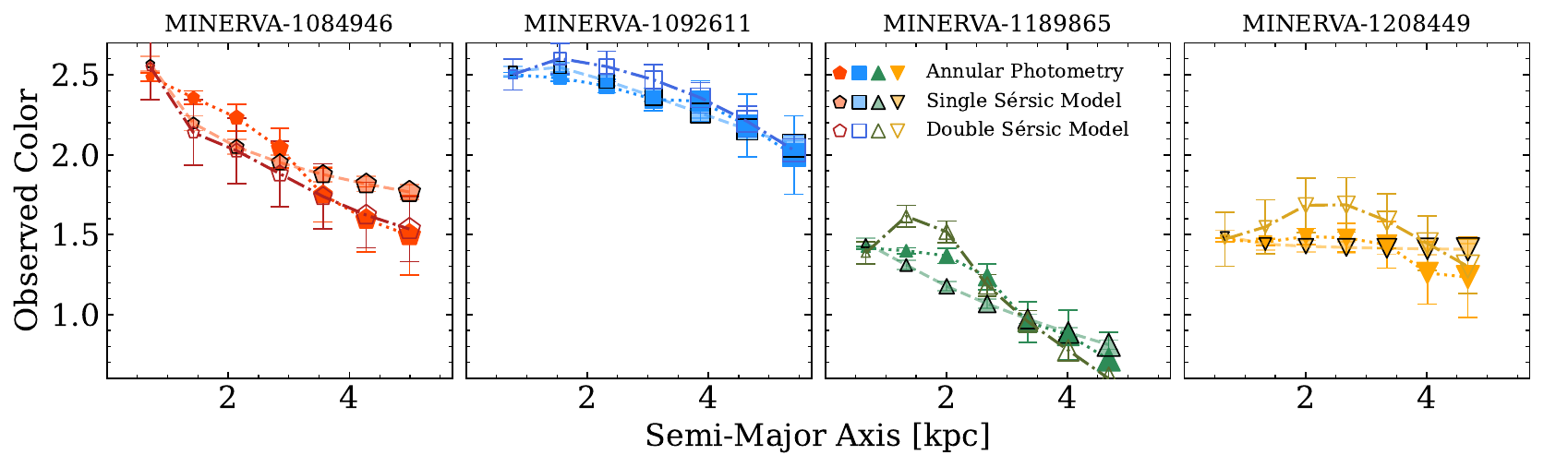}
    \caption{Observed color gradients measured from aperture photometry are consistent with those extracted via forward-modeling techniques. Solid points show the color gradients derived from PSF-matched elliptical annuli, as shown in the top row of Figure \ref{fig:uvj}. The filters used to compute the color gradients are also identical to those in Figure \ref{fig:uvj}. The color gradients derived from single S\'ersic profile fits using \textsc{Pysersic} \citep{Pasha2023} are shown with black-outlined points, while open points show gradients measured from a double S\'ersic model.}
    \label{fig:gradcomp}
\end{figure*}

As mentioned in Section \ref{sec:annuli}, we do not correct our annular photometry profiles for the effects of PSF smoothing. One common way to account for these effects is by forward-modeling the photometry in each band through a best-fit 2D S\'ersic model \citep[e.g.,][]{Szomoru2010,Szomoru2012,Szomoru2013,Suess2019a,Martorano2026}. We perform a similar model-based analysis by independently fitting both one- and two-component S\'ersic models to each galaxy in all filters using \textsc{Pysersic} \citep{Pasha2023} and compare the resulting color gradients to those from our aperture photometry method in Figure \ref{fig:gradcomp}.

In general, Figure \ref{fig:gradcomp} shows that the color gradients are consistent within the $1\sigma$ uncertainties for all galaxies regardless of the method used. Notably, MINERVA-1189865 has significant differences in the observed color gradients at $R<2$ kpc (\arc{0.3}), with the double S\'ersic profile (open points) finding redder colors compared to the single S\'ersic profile (black-outlined points). The different findings of these two models highlight the assumptions made when modeling light profiles (i.e., two- or one-component S\'ersic models) are crucial. We also note that the aperture color gradient (solid points) appears to find a color gradient between the two models for MINERVA-1189865.

Overall, we opt to use the aperture photometry method presented in Section \ref{sec:annuli}, as the color gradients measured using all three methods are comparable and this method introduces the fewest assumptions. In particular, the forward-modeling methods require assumptions regarding the number of components used (i.e., single or double S\'ersic models), whether to model each filter independently or impose a global prior based on a, e.g., long-wavelength stack, how to incorporate residual flux into the color gradients, and if the galaxies are even well-described by a S\'ersic light profile. Moreover, by using an aperture photometry analysis, we can more directly highlight the impact of using a smaller photometric aperture when studying high-redshift, extended sources.

\section{Fiducial Model Dust and Metallicity Priors} \label{app:priors}
\setcounter{figure}{0}
\setcounter{table}{0}

\begin{figure}[b!]
    \centering
    \includegraphics[width=\linewidth]{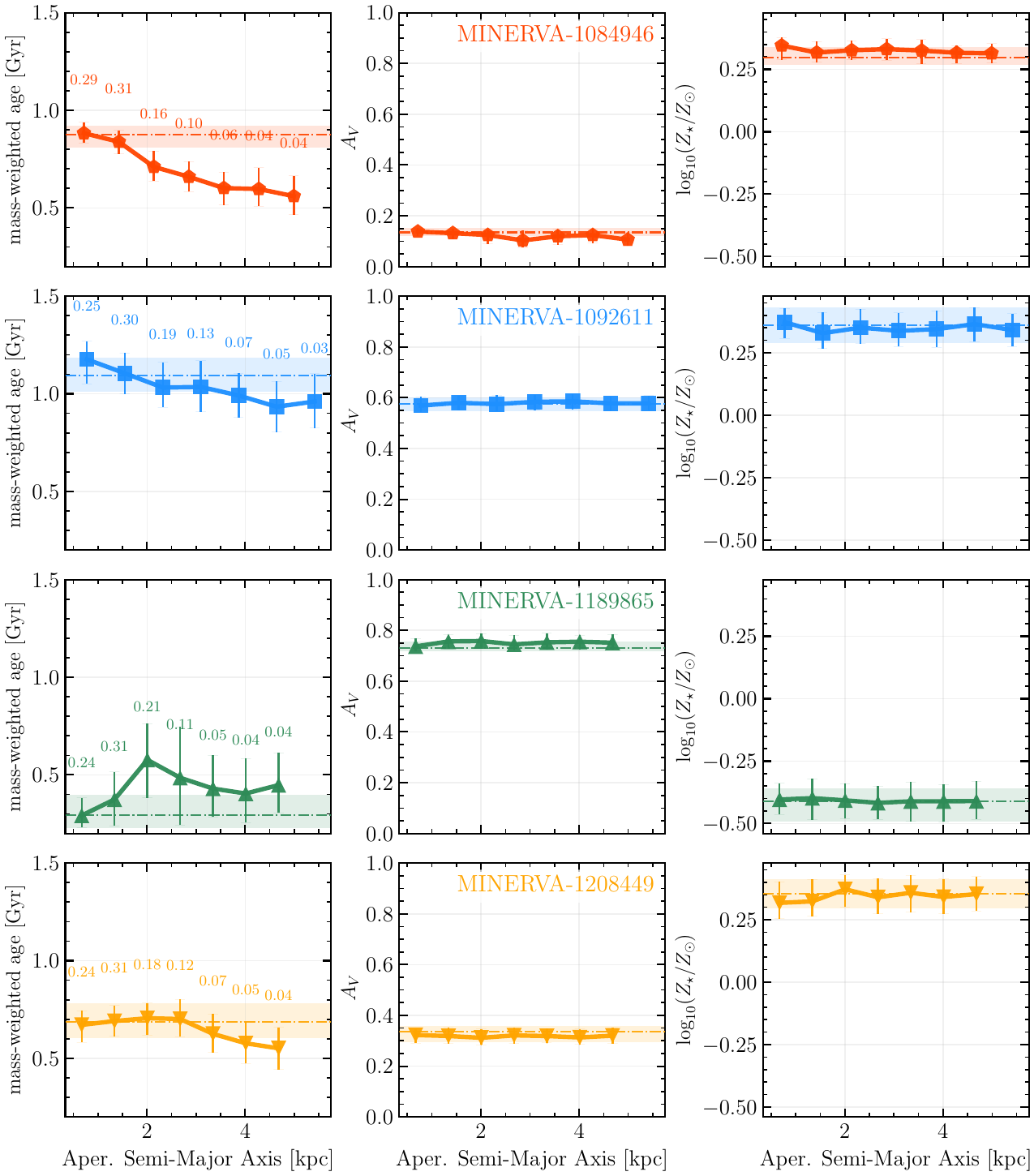}
    \caption{The distribution of galaxy mass-weighted age, $A_V$ and $\log(Z/Z_{\odot})$ with respect to the size of the semi-major axis of our annuli photometry for our ``fiducial'' model, demonstrating the effect of the strong metallicity Gaussian prior set by the values from \citetalias{Carnall2024}, and the strong dust prior set via modeling including MIRI photometry. Colors, lines, and symbols are identical to Figure \ref{fig:params}.}
    \label{fig:priors_fig}
\end{figure}

Our fiducial \pros{} model (for which we report our priors in Table \ref{table:fiducual}) includes strong priors on stellar metallicity ($\log(Z/Z_{\odot})$) from \citetalias{Carnall2024}, and on $\tau_V$ based on separate modeling of the global super ($D=\arc{1.4}$) photometry, also including MIRI photometry where available. In particular, for the metallicity we use a Gaussian prior on the $\log(Z/Z_{\odot})$ with $\mu$ equal to the median value from \citetalias{Carnall2024} and $\sigma$ equal to the average of the upper and lower limits from \citetalias{Carnall2024}. For the dust, we use a truncated Gaussian prior on the ``dust2'' optical depth (defined in FSPS as the dust optical depth at rest-frame 5500 \AA{} applied to stellar light older than 10 Myr) with $\mu$ equal to the median of our super+MIRI global fits, a $\sigma$ equal to 0.05 mag, and we truncate our prior at $\mu \pm \sigma$. With these priors applied, the radial changes in the $A_V$ (which is estimated assuming $A_{\rm V} = 1.086 \times $ dust2) and $\log(Z/Z_{\odot})$ are very small as compared to the radial change in the mass-weighted age of our four massive quiescent galaxies, as indicated in Figure \ref{fig:priors_fig}. This is to say that under the assumption of negligible gradients in stellar metallicity and dust attenuation, our objects are well-modeled with strong gradients in their mass-weighted ages. 

\renewcommand{\arraystretch}{0.9}
\begin{table}[h!]
\begin{threeparttable}
    \centering
    \footnotesize
    \caption{\pros{} Fitting Priors for Free Parameters in Our Fiducial Model.}
    \begin{tabular*}{\linewidth}{@{\extracolsep{\fill}}lcccr@{}}
        \hline\hline
        Parameter & $N_{\rm free}$ & Prior & ID & Values \\
        \hline
        redshift & 1 & Uniform & all & $z_{\rm spec}$ $\pm$ 0.0015 \\
        $\log(M_{\rm form}/M_{\odot})$ & 1 & Uniform & all & [6.0, 12.5] \\
        $\log(Z/Z_\odot)$ & 1 & Normal & 1092611 & (0.35, 0.075) \\
        ... & ... & ... & 1084946 & (0.32, 0.045) \\
        ... & ... & ... & 1189865 & ($-0.41$, 0.075) \\
        ... & ... & ... & 1208449 & (0.35, 0.070) \\
        
        log(${\rm SFR_{ratios}}$) & 10 & Student's-t & all & (0, 0.3) \\
        dust2 & 1 & Truncated Normal & 1092611 & (0.534, 0.046) \\
        ... & ... &  ... & 1084946 & (0.101, 0.046) \\
        ... & ... &  ... & 1189865 & (0.700, 0.046) \\
        ... & ... &  ... & 1208449 & (0.295, 0.046) \\
        dust1\_fraction & 1 & Uniform & all & [0, 2] \\
        
        dust\_index & 1 & Uniform & all & [$-$1.0, 0.4] \\
        
        $\log(Z_{\rm gas}/Z_\odot)$ & 1 & Uniform & all & [$-2.0$, 0.5] \\
        $\log U$ & 1 & Uniform & all & [$-4.0$, $-1.0$] \\
        $\log(f_{\rm AGN})$ & 1 & Uniform & all & [$-5.0$, 0.48] \\
        $\log(\tau_{\rm AGN})$ & 1 & Uniform & all & [0.7, 2.18] \\
        \hline
        \label{table:fiducual}
    \end{tabular*}
    \begin{tablenotes}
        \item\hspace{-2pt}\textbf{Note.} Values reported in square brackets represent uniform boundaries, while values reported in soft brackets represent object-specific mean and sigma sets used.
    \end{tablenotes}

\end{threeparttable}
\end{table}

\section{Effects of the $z=20$ Star Formation Limit} \label{app:zmax20}
\setcounter{figure}{0}
\setcounter{table}{0}
To assess the impact of imposing no star formation before $z=20$ in our \pros{} modeling, we fit the slit-based photometry with the same modeling assumptions as we use for our fiducial model with the strong prior on dust attenuation removed. Then, we reran the fitting with the single modification of removing the $z_{\rm max}=20$ limit on star formation used in our fiducial model. This change resulted in typically higher stellar mass assembly histories at $z\gtrsim8$ for all four of our objects. To test the impact of this modification on the goodness of fit, we estimated the Bayes' factor ($BF$) between the two models as the ratio of the maximum natural log-evidences (and their uncertainties) as determined by \textsc{dynesty} within the \pros{} fitting procedure:
\begin{equation}
\label{eq:BF}
    BF = \exp\left\{\ln(\mathcal{Z}_{{\rm z_{\rm max,SF}}=20}) - \ln(\mathcal{Z}_{{\rm z_{\rm max,SF}}=\infty})\right\}.
\end{equation}
We find an average factor of $BF\sim 1.50 \pm$0.15. This indicates weak/anecdotal evidence in favor of our modeling with $z_{\rm max,SF} = 20$, suggesting that the data do not prefer either of two models in a statistically significant way. Therefore, we conclude that prior/modeling choices such as setting the maximum redshift of star formation can have significant effects on the SFHs of our galaxies at early times without significant constraints from data, and therefore such choices must be made carefully.

\section{Super+MIRI \pros{} modeling} \label{app:miri-seds}

\begin{figure*}[p!]
    \centering
    \includegraphics[width=\linewidth]{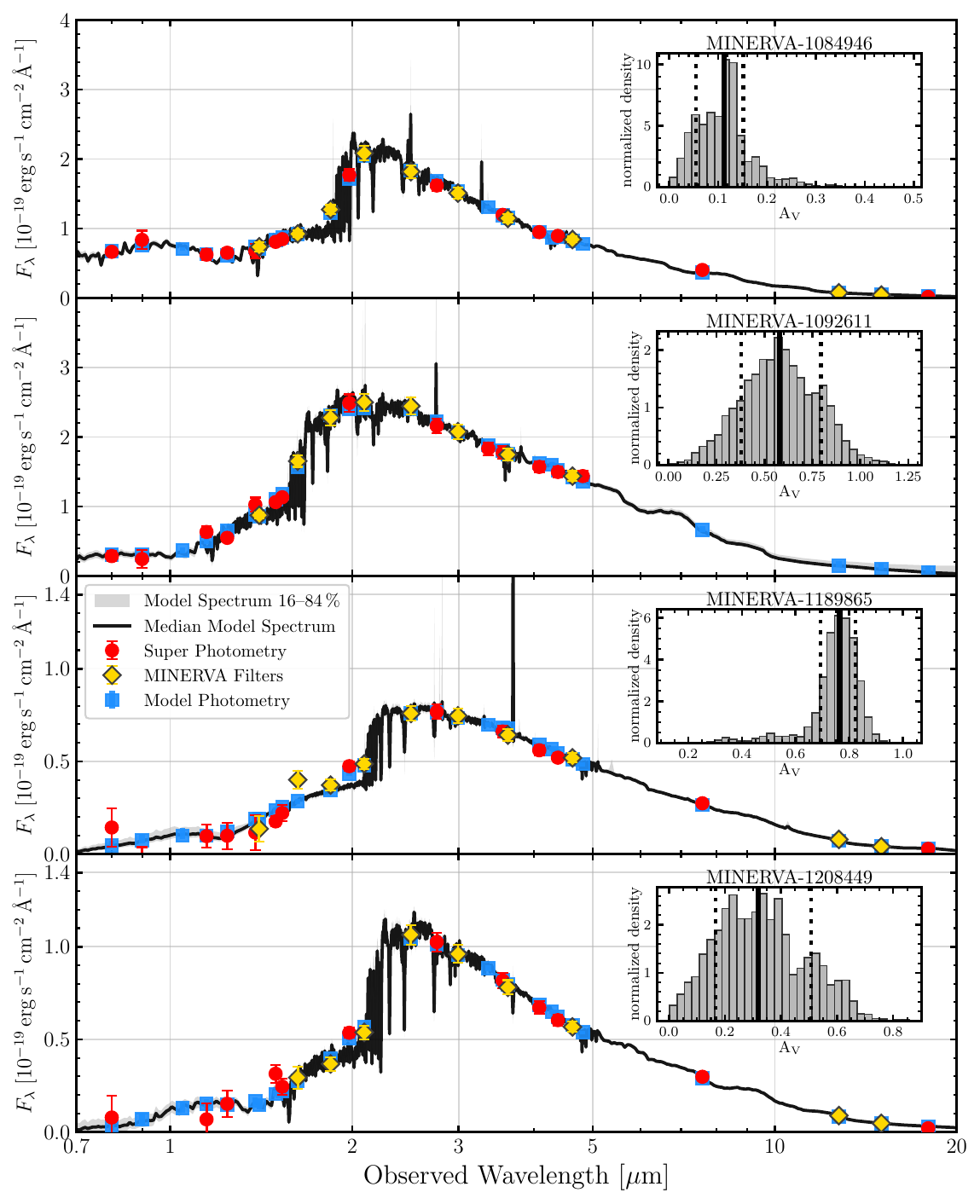}
    \caption{Posterior SEDs for each of our ultramassive quiescent galaxies using our \pros{} modeling of the super and MIRI combined photometry. Median model spectra and photometry, as well as their respective 1$\sigma$ uncertainties, are shown. The observed data are shown in red. Inset are the posterior distributions of $A_{\rm V}$ for each object. The median of each distribution is shown as the vertical solid black line, and the upper and lower 1$\sigma$ uncertainties are shown as dotted black vertical lines. }
    \label{fig:pros-wmiri}
\end{figure*}

Modeling of the combined super and MIRI photometry (where available) in F770W, F1280W, F1500W, and F1800W are used in tandem with the super ($D=\arc{1.4}$) color aperture photometry to measure a global value of $A_{\rm V}$. We use the medians of each galaxy's posterior distribution (which are reported in Table \ref{table:fiducual}) in $A_{\rm V}$ as the central values of the truncated Gaussian prior in our fiducial modeling. The \pros{} best-fit SED and model spectra are shown in Figure \ref{fig:pros-wmiri}. Insets show the posterior distribution of $A_V$ for each source.

Although MINERVA-1092611 does not have coverage in the MIRI filters, for consistency in our methodology we still measure the $A_{\rm V}$ and apply it to our fiducial model in the same manner as the other three galaxies using the super photometry. As a check, we compared the $A_{\rm V}$ posterior distributions for MINERVA-1092611 using the modeling of the super photometry with modeling of the innermost annulus of MINERVA-1092611 with the strong dust prior removed (that is, so that both models have identical priors). We find that the median values of both posteriors agree well within 1$\sigma$. See Appendix \ref{app:priors} for more information on the specific priors we use in our fiducial model. 

\clearpage
\bibliographystyle{aasjournal}
\bibliography{references}
\end{document}